# Scalar and fully relativistic pressure and temperature-dependent *ab-initio* thermodynamics study of simple cubic polonium

Balaram Thakur[1*], Xuejun Gong[1,2], and Andrea Dal Corso[1,3]

[1]International School for Advanced Studies (SISSA), Via Bonomea 265, 34136 Trieste, Italy.
[2]School of Physical Science and Technology, Xinjiang University, Urumqi, Xinjiang, 830046, China.
[3]CNR-IOM, Via Bonomea 265, 34136 Trieste, Italy.

*Corresponding author: bmtr789@gmail.com

**Abstract:**

The ab-initio thermodynamic properties of simple cubic polonium (α-Po) were studied within the quasi-harmonic approximation (QHA), where both lattice vibrations (phonons) and electronic excitations contributions are included in the Helmholtz free energy. We investigate the influence of spin-orbit coupling (SOC) by comparing the scalar relativistic (SR) and fully relativistic (FR) pseudopotentials on the thermodynamic properties of polonium and evaluate the performance of three popular exchange-correlation functionals, GGA (PBE and PBEsol), and LDA (PZ). Temperature and pressure-dependent thermodynamic properties were compared with the available experimental and theoretical studies. We found that the effect of electronic excitations is negligible for all the thermodynamic properties. LDA+SOC provides a better agreement with the experimental volume, while the thermal expansion coefficients from LDA+SOC and PBEsol + SOC closely match experimental values. SOC effects appear insignificant for the isobaric heat capacity but substantially contribute to the adiabatic bulk modulus. The phonon dispersions and mode-Grüneisen parameters ($\gamma_{\mathbf{q}\eta}$) were interpolated at the lattice constant corresponding to 301 K. The SR dispersions exhibit several anomalies in all directions, which were suppressed significantly by the inclusion of SOC. The role of SOC on the elastic constant-coefficient ($C_{ij}$) and elastic anisotropy factor at 0 K is also studied. The Pugh ratio confirmed that simple cubic polonium is ductile, and we observed that the elastic anisotropy factor and Pugh ratio decreased with increasing pressure. The effect of SOC on the Cauchy pressure is also determined.

Email:  Balaram Thakur (bmtr789@gmail.com), Xuejun Gong (xgong@sissa.it), Andrea Dal Corso (dalcorso@sissa.it)

**Introduction:**

Polonium, a member of the chalcogen family (group 16), is the only element in the periodic table that crystallizes in a simple cubic structure (SC) at low temperatures and ambient pressure. Other elements like phosphorus, calcium, and arsenic also exist in a simple cubic structure, but only at pressures above 10 GPa, 32 GPa, and 25 GPa, respectively [1]. Polonium-209 and its other known isotopes are radioactive, and its isotope Po-210, as a strong α emitter, has found applications in thermoelectric generators in space technology, as a neutron source in Po-Be alloy.

Despite its scientific significance and technological utility, the experimental investigation of polonium remains limited, primarily due to its high radioactivity, toxicity, and scarcity. The experimental values of a few physical properties like melting point (527 K), boiling point (1235 K), vapour pressures, electrical resistivity, density at 298 K, linear thermal expansion coefficient at 298 K, and thermal conductivity at 300 K are documented[2]. Theoretically, most of the attention is drawn towards understanding the existence in the simple cubic phase, which contrasts with the isoelectronic elements like selenium and tellurium existing in the rhombohedral (trigonal) phase. Polonium has two metallic phases: α-Po (SC) and β-Po (rhombohedral), the latter being a stretched version of α-Po along the [111] direction, with a transition at 348 K [3]. Also, at low pressure (~ 2 GPa), the simple cubic undergoes a structural phase transition into a mixture of two trigonal (86° and 97.9°) phases [4].

The role of relativistic effects and spin-orbit coupling (SOC) is crucial for revealing the underlying bonding mechanisms, structural stability and predicting physical properties such as superconductivity, which has been theoretically suggested with a critical temperature $T_C \sim 4$ K and is due to the influence of SOC on the transverse soft phonon mode at $\mathbf{q} \approx 2/3$ **R** [5]. Phonon dispersions of SC polonium exhibit Kohn anomalies arising from Fermi surface nesting, which are significantly reduced when SOC is included in the calculations [4–7]. This softening of the phonon modes due to SOC contrasts with that phonon dispersions observed in other high-$Z$ metals such as lead [8,9].

The elastic constant coefficients determined with different methods can be found in Refs.[3,10,11]. The α-Po exhibits pronounced elastic anisotropy due to the emptiness of the simple cubic unit cell along the body diagonal, promoting easier deformation in the body-diagonal direction compared to along the cube edges [10,11]. The positive Poisson ratio measured in all crystallographic directions confirms that α-Po contracts laterally when stretched longitudinally [11]. Furthermore, analysis based on Cauchy pressures and Pugh and Frantsevich criteria confirms the metallic and ductile nature of SC polonium [11].

Although earlier studies have examined the structural stability of polonium, its temperature and pressure-dependent thermodynamic behaviour, understanding the role of SOC and their behavior for different *xc* functionals has not been systematically investigated. For instance, Verstraete[6] uses the density functional perturbation theory (DFPT) and quasi-harmonic approximation (QHA) and reports the temperature-dependent thermal expansion and isobaric heat capacity only at 298 K. Verstraete [6] used only LDA and studied both SR and FR pseudopotentials. In the present paper, we systematically analysed the role of SOC with different *xc* functionals on the pressure and temperature-dependent bulk modulus, isobaric heat capacity, volume thermal expansion coefficient, and thermodynamic average Grüneisen parameter of α-polonium.

**Theory and Methods:**

The thermodynamic and mechanical properties of polonium were calculated using the density functional theory (DFT) within quasi-harmonic approximation (QHA), as implemented in the `Thermo_pw`[12] (v2.0.0) code, a `FORTRAN` driver of the Quantum ESPRESSO (QEv7.3) routines [13,14]. Both scalar relativistic (SR) and fully relativistic (FR) ultrasoft (US) [15] PPs (from *pslibrary1.0.0* [16,17]) were employed, and they were generated according to a modified Rappe-Rabe-Kaxiras-Joannopoulos ultrasoft (RRKJ) [18] scheme. In FR PPs, the SOC is included, as discussed in detail in Refs.[19,20] for ultrasoft PPs (USPP) [15].

For each type of PP (SR and FR), we considered three exchange-correlation functionals approximated by the local-density approximations (LDA) with the Perdew-Zunger (LDA-PZ) [21] parameterization, the generalized gradient approximation (GGA) suggested by Perdew-Burke-Ernzerhof (PBE) [22], and the PBE functional modified for densely packed solids (PBEsol) [23]. Specifically, for LDA, PBE, and PBEsol, we used `Po.pz-dn-rrkjus_psl.1.0.0.UPF`, `Po.pbe-dn-rrkjus_psl.1.0.0.UPF`, and `Po.pbesol-dn-rrkjus_psl.1.0.0.UPF` files for SR PPs, while `Po.rel-pz-dn-rrkjus_psl.1.0.0.UPF`, `Po.rel-pbe-dn-rrkjus_psl.1.0.0.UPF`, and `Po.rel-pbesol-dn-rrkjus_psl.1.0.0.UPF` files for FR PPs, from *pslibrary* [16,17]. The electronic configuration of polonium is [Xe] $4f^{14}6s^{2}6p^{4}5d^{10}$, and in the above PPs, the 5$d$, 6$s$, and 6$p$ were treated as valence states, while the 4$f$ states are frozen in the core and are accounted for by the non-linear core correction [24].

The pseudo-wavefunctions and charge densities were expanded in a plane wave basis set with a kinetic energy cutoff of 80 Ry and 400 Ry, respectively. Brillouin zone integrations were performed on a uniform 32 × 32 × 32 **k**-point mesh using the Monkhorst-Pack method [25]. The Methfessel and Paxton (MP) smearing approach [26] with a MP smearing parameter σ = 0.02 Ry is used to deal with the Fermi surface. Phonon calculations were carried out using Density functional perturbation theory (DFPT) [27] extended to US-PP [20,28]. These parameters were determined through extensive convergence tests (see supplementary data S1), ensuring that the total energies, phonon frequencies, and elastic constant coefficients were converged within 1mRy, 1 cm$^{-1}$, and 1 GPa, respectively.

The dynamical matrices were calculated on a coarse 8 × 8 × 8 **q**-point grid for each pseudopotential and functional and then interpolated onto a thicker 192 × 192 × 192 **q**-point grid using the Fourier interpolation method. For QHA, we calculated the dynamical matrices on different geometries with a step size of 0.025 a.u. For the FR case with LDA and PBEsol, we used 6 geometries, whereas for all other cases, we used 7 geometries. More details for the geometries used in each case are given in the supplementary data S2. These ranges were chosen to ensure real (positive) phonon frequencies and satisfaction of Born stability criteria across all geometries.

The relations used to calculate the thermodynamic and mechanical properties are documented in the `Thermo_pw`[12] manuals and in Refs.[29–35]. The main goal is to determine the Helmholtz free energy F (V, T) using direct summation over phonon frequencies and is read as:

$$F(V, T) = U_0(V) + F_{ph}(V,T) + F_{elec}(V,T) \quad (1)$$

In equation (1), $U_0(V)$ is the static (0 K) total energy of a system obtained from DFT, $F_{ph}(V,T)$ represents the lattice vibrations (phonons) contribution to the free energy and $F_{elec}(V,T)$ is the electronic contribution to the free energy. For clarity and completeness, we summarised the key theoretical concepts used in this study in the supplementary data S3.

The elastic constant coefficients ($C_{ij}$) were determined from the stress tensor components induced by a strain on the equilibrium geometry at 0 K. Furthermore, using these single crystal $C_{ij}$'s values, tetragonal shear moduli $C'=(C_{11}-C_{12})/2$ and elastic anisotropy factor $A=C_{44}/C'$ are obtained. Finally, the ductile or brittle nature of polonium is evaluated using the Pugh ratio (G/B), where G and B are the polycrystalline shear (G) and bulk moduli (B) calculated using $C_{ij}$'s and the Voigt-Reuss-Hill (VRH) averaging relations [36].

**Results and Discussion:**

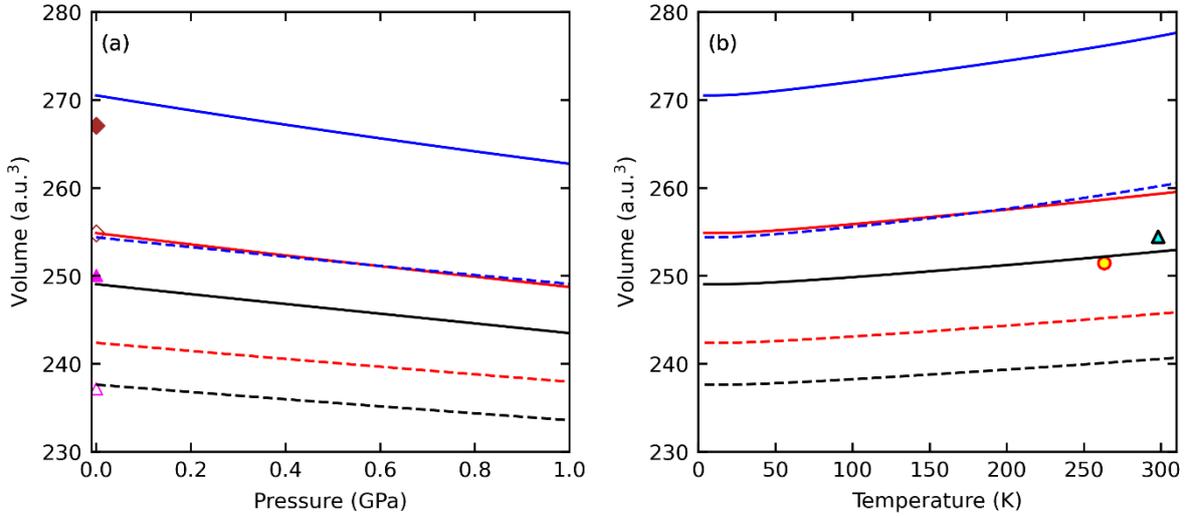

**Fig. 1:** Volume variation with (a) pressure at 4 K and (b) temperature at 0 kbar. Solid and dotted lines represent the fully relativistic and scalar relativistic cases, respectively. For each case, black, blue, and red represent the LDA, PBE, and PBEsol. In (a), results from all-electron calculations [10] are included; diamonds represent GGA+SOC (filled) and GGA (empty); triangles denote LDA+SOC (filled) and LDA (empty). In (b), the experimental volume at 263 K (circle) [37] and 298 K (square) (calculated from the density 9.2 g/cm$^3$) [2] are compared.

Fig.1(a) shows the equation of state (EOS) at 4 K for LDA (black), PBE (blue), and PBEsol (red), with (solid lines) and without (dashed lines) SOC. The EOSs are typically exponential in form, but here, since the pressure range is very small, the EOS in Fig.1(a) appears nearly linear. The EOS is fitted with the 4$^{th}$-order Birch-Murnaghan equation, and the equilibrium parameters like the lattice constant, bulk modulus ($B_T$), and pressure derivative of the bulk modulus ($B'_T$) are summarised in Table 1 and compared with the available theoretical and experimental reports. Table 1 shows that including SOC increases the volume by ~ 6.3 % in PBE and ~5 % for both PBEsol and LDA. These agree with the all-electron results [10], where the increase in the equilibrium volume is observed for both LDA and GGA by about 4-7 %. In contrast, a semi-empirical tight-binding (TB) study showed only a 2.5 % increase upon including the SOC [38]. Our lattice constant for PBE and LDA, both with and without SOC, agrees with the US-PP study of Belabbes et al. [7]. Moreover, we found that with SOC, the bulk modulus decreases significantly by around 29 % for PBE, 25 % for PBEsol, and 24 % for LDA. This decrease in bulk modulus is correlated with the enhancement in the lattice constant (discussed above). For both with and without SOC, our $B_T$ for LDA are close and PBE results are lower when compared with the all-electron results [10], whereas only LDA+SOC and PBE are in agreement with Belabbes et al. [7].

Fig. 1(b) illustrates the temperature dependence of the equilibrium volume. Comparing with the experimental volumes at 263 K and 298 K, we found that our LDA+SOC result is in closest agreement. The change in volume with the introduction of SOC at 0 kbar remains nearly constant with temperature and is about 5 % for LDA and PBEsol cases and ~ 6 % for PBE, as shown in Fig.1(b). This change in volume decreases slightly at higher pressure, for example, at 10 kbar (not shown here), and it is about ~ 4 % for LDA and PBEsol and ~ 5% for the PBE case.

|  |  | Lattice constant ($a$) (in a.u.) | Bulk modulus ($B_T$) (in GPa) | Pressure derivative of $B_T$ ($B'_T$) |
|---|---|---|---|---|
| This study (US-PP) | LDA | 6.19 | 56.8 | 4.73 |
|  | LDA+SOC | 6.29* | 42.5* | 5.37* |
|  | PBE | 6.33 | 45.8 | 5.96 |
|  | PBE+SOC | 6.46* | 31.3* | 6.16* |
|  | PBEsol | 6.23 | 52.4 | 4.71 |
|  | PBEsol+SOC | 6.33* | 39.4* | 5.04* |
| Other studies | LDA (FLAPW) [10] | 6.19 | 57.1 |  |
|  | LDA+SOC (FLAPW) [10] | 6.30* | 42.3* |  |
|  | LDA (US-PP) [7] | 6.20 | 52.6 |  |
|  | LDA+SOC (US-PP) [7] | 6.27* | 42.8* |  |
|  | LDA (NC-PP) [3] | 6.19 | 56.0 |  |
|  | LDA+SOC (HGH) [6] | 6.30* | 39.45* | 4.89* |
|  | GGA (FLAPW) [10] | 6.34 | 46.9 |  |
|  | GGA+SOC (FLAPW) [10] | 6.44* | 40.1* |  |
|  | PBE (US-PP) [7] | 6.31 | 45.6 |  |
|  | PBE+SOC (US-PP) [7] | 6.44* | 35.0* |  |
|  | HL (TB) [38] | 6.16 | 59.0 |  |
|  | HL+SOC (TB) [38] | 6.21* | 51.0* |  |
|  | HL (FPLAPW) [38] | 6.31 | 44.0 |  |
| Experiments |  | 6.312 (263 K)[37], 6.348[39], 6.337 (298 K)[2] |  |  |

**Table 1:** Equilibrium structural parameters: unit cell lattice constant ($a$), bulk modulus ($B_T$), and pressure derivative of bulk modulus ($B'_T$) at 4 K for LDA, PBE, and PBEsol. The values with and without * represent calculations with and without SOC. The effect of zero-point energy is included in free energy. The theoretical result of Legut *et al.* [10] using the all-electron full-potential linearised augmented plane waves method (FLAPW) and Belabbes *et al.* [7] using US-PP. The value of scalar relativistic norm-conserving (NC)PP [3] is included. The results from Ref.[6], where pseudopotential methods of the HGH form and LDA+SOC, and from Ref.[38], where the values using FPLAPW and NRL tight-binding (TB) method with *xc* of Hedin–Lundqvist (HL) prescription are used.

Fig.2(a) and Fig.2(b) illustrate the volume thermal expansion (β) as a function of temperature at 0 kbar and as a function of pressure at 301 K, respectively. Fig.2(a) reveals that the electronic excitations contribution in β is negligible for all the cases. We notice that the experimental β values agree with our LDA+SOC and PBEsol + SOC results. In contrast, theoretical data from Ref.[6], obtained with LDA+SOC, predict values less than ours. It is noted that the relativistic corrections in Ref.[6] are included through a Hartwigsen-Goedecker-Hutter (HGH) pseudopotential, and a significant negative thermal expansion coefficient was noticed at very low temperatures. In contrast, our results show a slightly negative β at very low temperatures, but its magnitude is negligible. Furthermore, Fig.2(a) reveals a difference in the β values obtained with and without SOC across the entire temperature range and is independent of the functionals used. At 301 K and 0 kbar, the SOC increases the β by approximately 24.2 % for LDA, 24.3 % for PBEsol, and only 7 % for PBE, whereas at 10 kbar it is about 18.3 %, 32.7 %, and 11.9 %, respectively, as illustrated in Fig.2(b).

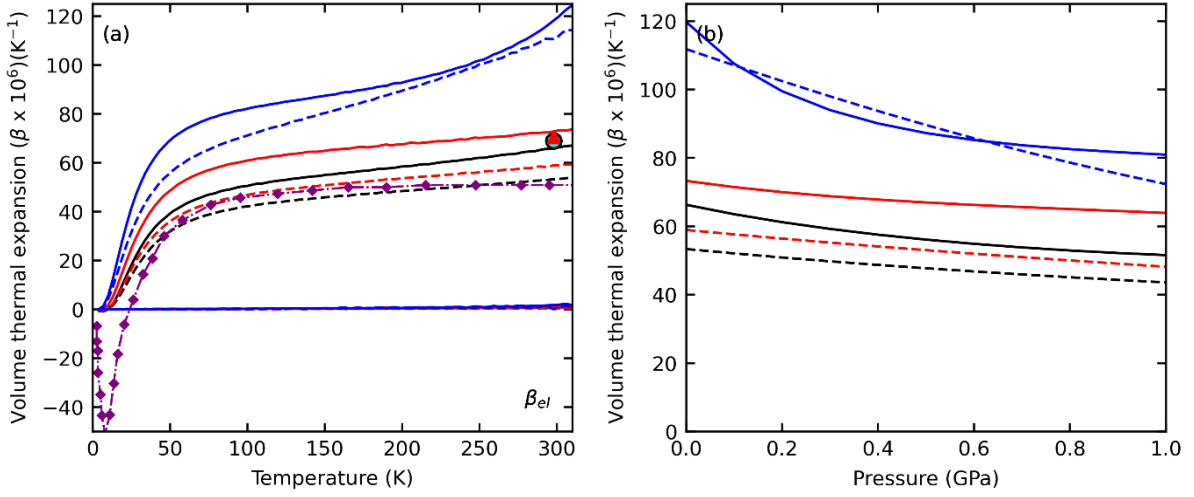

**Fig. 2:** (a) Temperature and (b) Pressure-dependent coefficient of volume thermal expansion at 0 kbar and 301 K, respectively. The solid and dotted lines represent the fully relativistic and scalar relativistic cases, and the blue, red, and black represent the PBE, PBEsol, and LDA. The circle [40] and triangle [2] symbols are experimental results, and the purple line + diamond is the LDA+SOC results from the theoretical study [6].

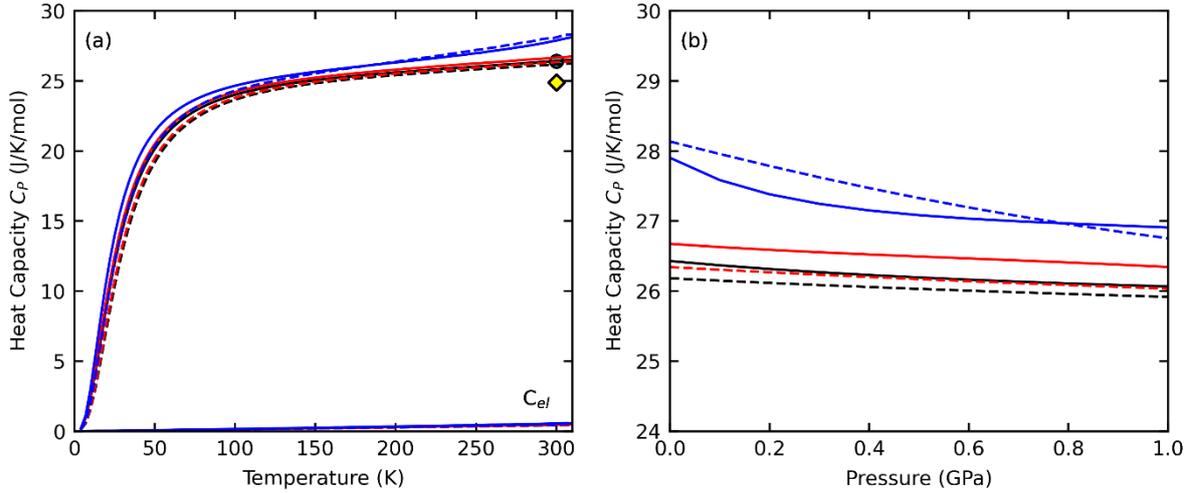

**Fig. 3:** (a) Temperature and (b) Pressure-dependent isobaric heat capacity at 0 kbar and 301 K, respectively. The solid and dotted lines represent the fully relativistic and scalar relativistic cases, and the blue, red, and black represent the PBE, PBEsol, and LDA. The circle is the experimental result from Ref.[41], and the diamond is from the LDA+SOC result [6].

Fig. 3(a) shows the temperature dependence of isobaric heat capacity ($C_P$). Comparing our results with experiment at 298 K [41], we found that LDA and PBE agree well for both with and without SOC cases. However, our LDA+SOC values are higher than those reported in Ref.[6]. The electronic excitation contribution in the isobaric heat capacity $C_{el}$ is negligible (see Fig.3(a)). Furthermore, in contrast to the β, the effect of SOC on $C_P$ is insignificant. In particular, LDA and PBEsol with and without SOC are almost the same. The higher values of $C_P$ for PBE, both with and without SOC, with respect to PBEsol and LDA, are due to different β, as shown in Equation

4. Moreover, Fig.3(b) reveals that at 301 K, the influence of SOC on $C_P$ is negligible (~ 1%) for all the functionals, even when the pressure is increased from 0 GPa to 1 GPa.

Fig. 4(a) and Fig.4(b) show the temperature dependence of isobaric ($B_T$) and adiabatic ($B_S$) bulk moduli calculated with and without SOC. For comparison, the all-electron[10] and US-PP[7] bulk moduli at 0 K (also shown in Table 1) calculated for LDA and PBE, both with and without SOC, are shown. As seen in Fig.4, at 0 K, our SR LDA is in good agreement whereas PBE is lower than the reported values. Moreover, our LDA+SOC is close to all-electron[10] and PBE+SOC is in good agreement with both all-electron[10] and US-PP[7] reports. Furthermore, we found that at 0 GPa and 301 K, the contribution of the SOC to adiabatic bulk modulus is around ~ 26 %, 28 %, and 23 % for LDA, PBE, and PBEsol, respectively.

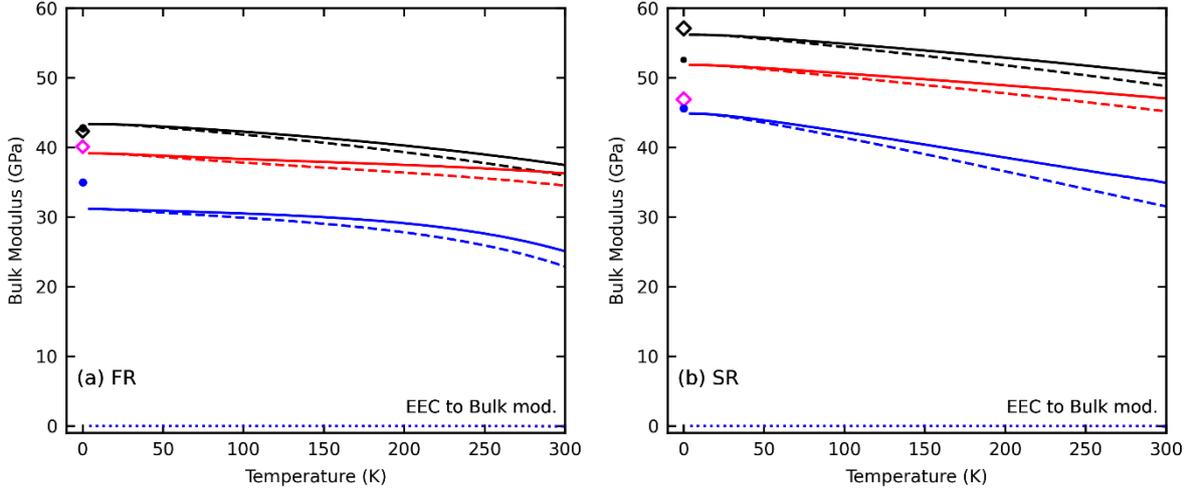

**Fig. 4:** Temperature-dependent isobaric ($B_T$) and adiabatic ($B_S$) bulk modulus for (a) fully and (b) scalar relativistic cases at 0 kbar. The black, blue, and red represent LDA, PBE, and PBEsol. The solid and dashed lines represent the $B_S$ and $B_T$. The dotted line corresponds to the electronic excitation contribution in the $B_S$. The symbols, empty diamond (magenta-PBE and black-LDA), are from all-electron [10] calculations, and filled circle (blue-PBE and black-LDA) are theoretical results from US-PP [7] methods. The symbols in (a) and (b) were calculated using with and without SOC, respectively.

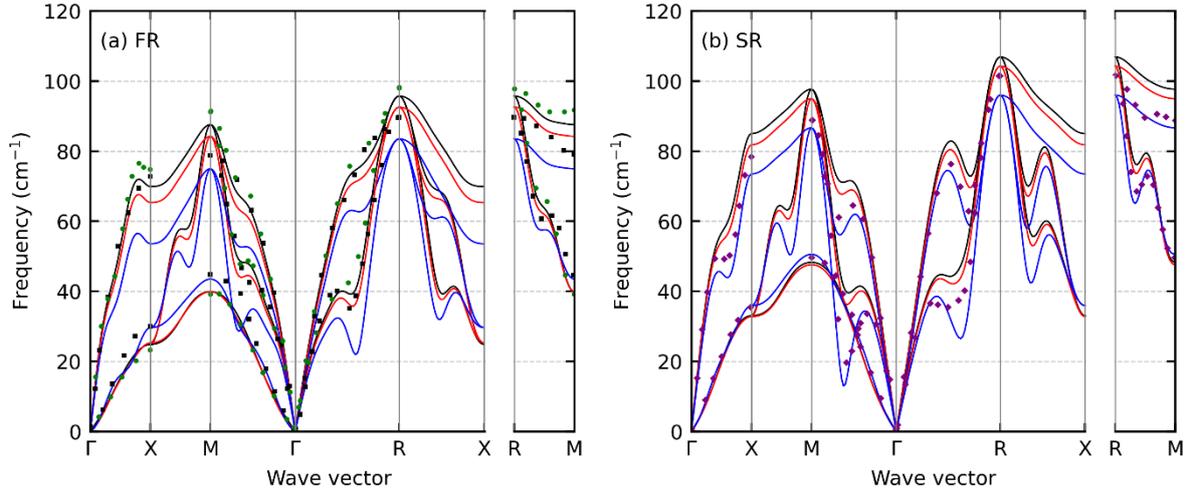

**Fig.5:** The phonon dispersion of polonium with (a) fully and (b) scalar relativistic case. The black, blue, and red lines are for LDA, PBE, and PBEsol, respectively. The phonon dispersions are interpolated on the lattice constant at 301 K (for FR- PBE: 6.521 a.u., PBEsol: 6.377 a.u., LDA: 6.323 a.u.; for SR-PBE: 6.385 a.u., PBEsol: 6.264 a.u., LDA: 6.219 a.u.). In (a), the square (black)[5] and circle (green)[6] are other LDA+SOC results, and in (b), the diamond (purple) is the LDA result from Ref.[6]. The frequencies at high-symmetry points are compared in Table 2.

| | Methods | | $\omega_L$ (**M**) | $\omega_T$ (**M**) | $\omega_{L,T}$ (**R**) | $\omega_L$ (**X**) | $\omega_T$ (**X**) |
|---|---|---|---|---|---|---|---|
| This study (US-PP) | LDA | 6.219 a.u. | 97.7 | 48.4 | 106.9 | 84.9 | 32.8 |
| | LDA + SOC | 6.323 a.u. | 87.6 | 39.5 | 95.9 | 69.9 | 24.5 |
| | PBE | 6.385 a.u. | 86.4 | 50.6 | 95.6 | 73.2 | 36.2 |
| | PBE + SOC | 6.521 a.u. | 74.4 | 43.2 | 83.3 | 53.6 | 29.4 |
| | PBEsol | 6.264 a.u. | 94.5 | 47.5 | 104.2 | 81.8 | 32.8 |
| | PBEsol + SOC | 6.377 a.u. | 84.3 | 39.5 | 92.5 | 65.3 | 24.5 |
| Other Studies | LDA | 6.347 a.u. [5] | 81.9 | 48.5 | 93.8 | 66.9 | 33.4 |
| | LDA + SOC | 6.347 a.u. [5] | 78.7 | 44.3 | 89.5 | 72.8 | 29.8 |
| | LDA + SOC | 6.274 a.u. [4,7] | 95.5 | 35.6 | 96.7 | 76.1 | 33.2 |
| | LDA | 6.302 a.u. [6] | 88.8 | 49.1 | 101.5 | 77.8 | 34.8 |
| | LDA + SOC | 6.302 a.u. [6] | 91.3 | 38.9 | 98.0 | 74.2 | 22.7 |

**Table 2:** Phonon frequencies ($\omega$ in cm$^{-1}$) at high symmetry points for LDA, PBE, and PBEsol functional calculated with and without SOC. The lattice constant at which the phonon frequencies are interpolated is mentioned. The phonon frequencies from other studies are also compared.

Fig.5(a) and Fig.5(b) show phonon dispersions with and without SOC, respectively, interpolated at the lattice constant at 301 K. Unlike other heavier elements like lead, where the anomalies are confined to specific directions [8,9], Fig.5 shows that polonium exhibits anomalies in phonon dispersion along several symmetry directions. Furthermore, with the introduction of the SOC, the anomalies were reduced considerably, again in contrast with the lead [9]. The reduction in the softening of phonon modes with SOC agrees with the previous reports [5–7]. Table 2 compares the phonon frequencies obtained at high symmetry points with the values obtained from the theoretical reports [5–7], which were only performed with LDA. From Fig.5 and Table 2, we noticed that introducing the SOC decreased our phonon frequencies. On the contrary, in Ref.[5], it was found that the phonon frequencies of the longitudinal mode at **X** increase when SOC is included. Similarly, in Ref.[6], the longitudinal frequencies at **M** increase with SOC but decrease at **X**. However, in Ref.[5] and in Ref.[6], the phonon frequencies were calculated at the experimental lattice constant (6.347 a.u.) and the theoretical equilibrium lattice constant (6.302 a.u.), respectively. We also computed the phonon dispersions at the experimental lattice constant 6.337 a.u. (298 K[2]) and found our results consistent with the observations in Ref.[6] (see supplementary S4). Also, we noticed that at the same lattice constant, the effect of SOC is insignificant for longitudinal modes but significant for the transverse mode at all high symmetry points, except at **R**.

To elucidate the microscopic origin of the SOC effect on the phonons, we examined the electronic band structures for LDA, PBEsol, and PBE, both with and without SOC. The band structure and density of states (DOS) for PBE, at experimental lattice constant, is shown in Fig.6, and for other functionals determined at both experimental and at equilibrium lattice constant are included as supplementary data S5. From Fig.6, we note that the inclusion of SOC lifts the degeneracy of the *p*-derived valence bands near the Fermi level, producing a noticeable band splitting and a reduction of the density of states at the Fermi energy $N(E_F)$. Upon inclusion of SOC, the 6*p* band splits near the Fermi energy, causing a substantial reduction in the density of states at $E_F$, about 25-30 % calculated for experimental lattice constant and ~ 24 % for all the functionals when determined at the equilibrium lattice constant (see Table S5(A) and S5(B) in supplementary data S5). This reduction in N($E_F$) weakens Fermi-surface nesting, as evident from Fig.S6 (see supplementary data S6).

In Fig. S6(A), the contour plot for the SR and FR Fermi surfaces in the (2 0 0) and (1 $\bar{1}$ 0) planes are compared. The distribution of |**q**| shown in Fig.S6(B) indicates that the presence of SOC reduces the frequency of nesting vectors along certain directions. Min *et al.*[42] and Belabbes *et al.*[7] reported the nesting vector for FR case with **q** = $\frac{2}{3}$ along **Γ - X**, **X - M**, **M - Γ**, **Γ - R**, and **R - M** directions. Our results show overall reasonable agreement with Refs.[7,42], with the exception of the **R - M** direction where the magnitude of q differs. This analysis demonstrates that the key features of the Fermi surface responsible for phonon anomalies persists even with SOC and this is consistent with previous studies[7].

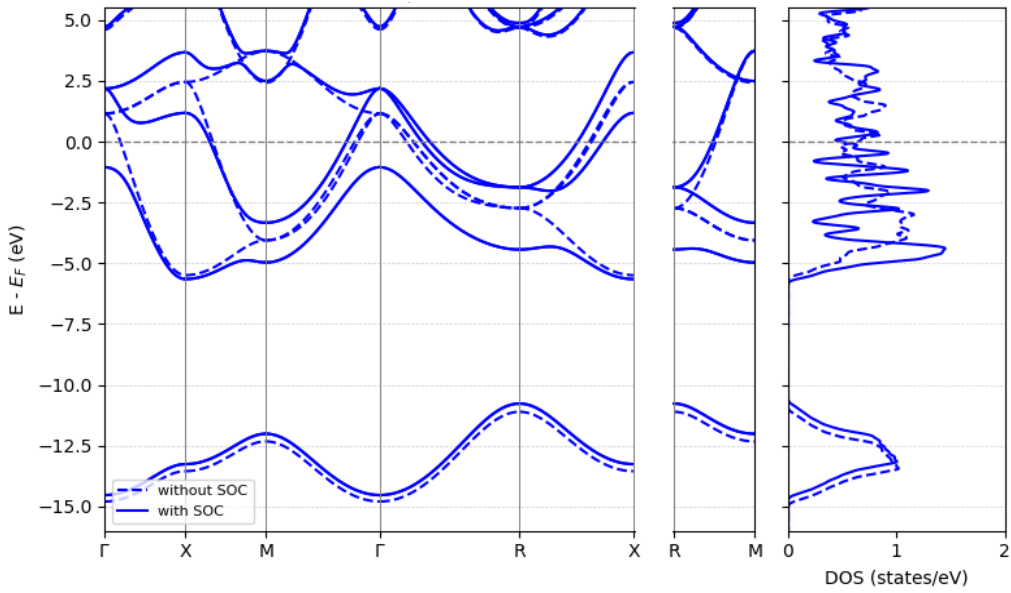

**Fig.6:** Band structure (left) and density of states DOS (right) for PBE calculated with and without spin-orbit coupling (SOC) at experimental lattice constant.

Fig.7(a) and Fig.7(b) show the mode-Grüneisen parameter ($\gamma_{\mathbf{q}\eta}$) obtained at 301 K for PBE, PBEsol, and LDA, with (FR) and without (SR) SOC, respectively. The inclusion of SOC leads to an overall decrease in $\gamma_{\mathbf{q}\eta}$, particularly for several modes along the [110] direction (**Γ** to **M**) and the [111] direction (**Γ** to **R**) for all the functionals. Using equation[43]:

$$\alpha = \frac{1}{B} \sum_{ks} \left(-\frac{\partial}{\partial V} \hbar \omega_{ks}\right) \frac{\partial}{\partial T}(n_{ks}) \qquad (2)$$

where the $n_{ks}$ is the Bose-Einstein occupation given as $n_{ks} = (e^{\frac{\hbar \omega_{ks}}{k_B T}} - 1)^{-1}$

We analysed the contribution of individual phonon modes to the thermal-expansion coefficient and found that modes with negative $\gamma_{\mathbf{q}\eta}$ correspond to the transverse acoustic modes (see supplementary data S7). Furthermore, from Fig. 7, we observe that $\gamma_{\mathbf{q}\eta}$ becomes negative along the **Γ-X** and **Γ-M**, primarily due to the softening of one of the transverse phonon branches. These negative $\gamma_{\mathbf{q}\eta}$ values indicate anomalous behavior, where the phonon frequencies of these transverse modes increase with lattice expansion. In contrast, along the **Γ-M** and **Γ-R**, the $\gamma_{\mathbf{q}\eta}$ curves display smooth maxima rather than sharp cusps, confirming that the anharmonic response remains continuous and no non-analytic behavior is present, except at **Γ**. These maxima originate mainly from the transverse modes and reflect a strong but gradual coupling between lattice strain and phonon frequencies.

Fig.8(a) and Fig.8(b) illustrate the temperature and pressure dependence of the thermodynamic average Grüneisen parameter (γ), respectively. As shown in Fig.8(a), the γ calculated without SOC is generally higher than the result calculated with SOC, except for the PBEsol case. This trend persists when the pressure increases from 0 to 10 kbar, as shown in Fig.8(b) at 301 K. The change in γ with temperature at 0 kbar and pressure at 301 K is only significant for the PBE case, while for LDA and PBEsol, the change remains minimal.

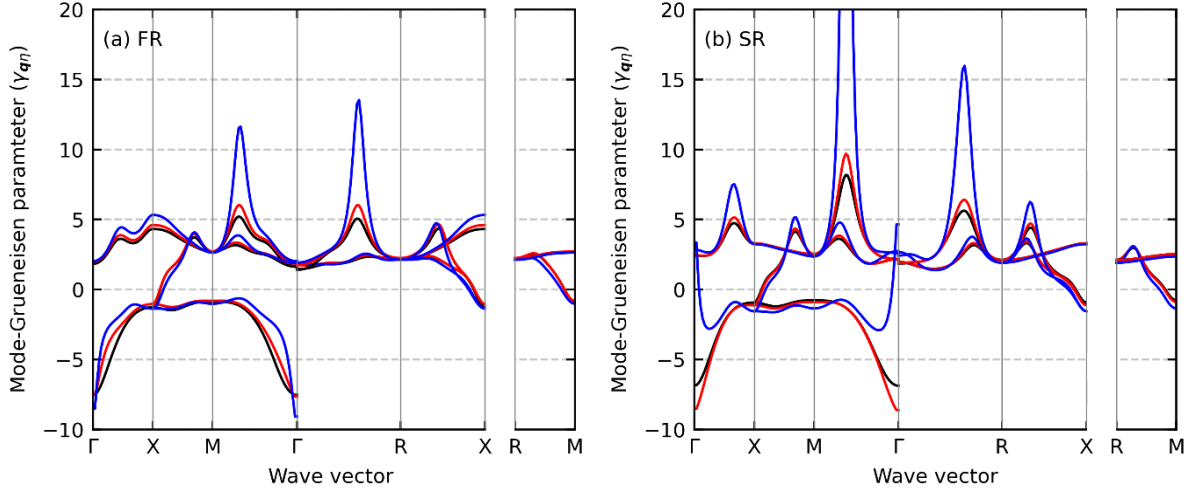

**Fig.7:** The Mode-Grüneisen parameter ($\gamma_{q\eta}$) with (a) fully and (b) scalar relativistic case. The black, blue, and red lines are for LDA, PBE, and PBEsol, respectively. They are calculated at the same lattice constant used for the phonon dispersions at 301 K (shown in Fig.5).

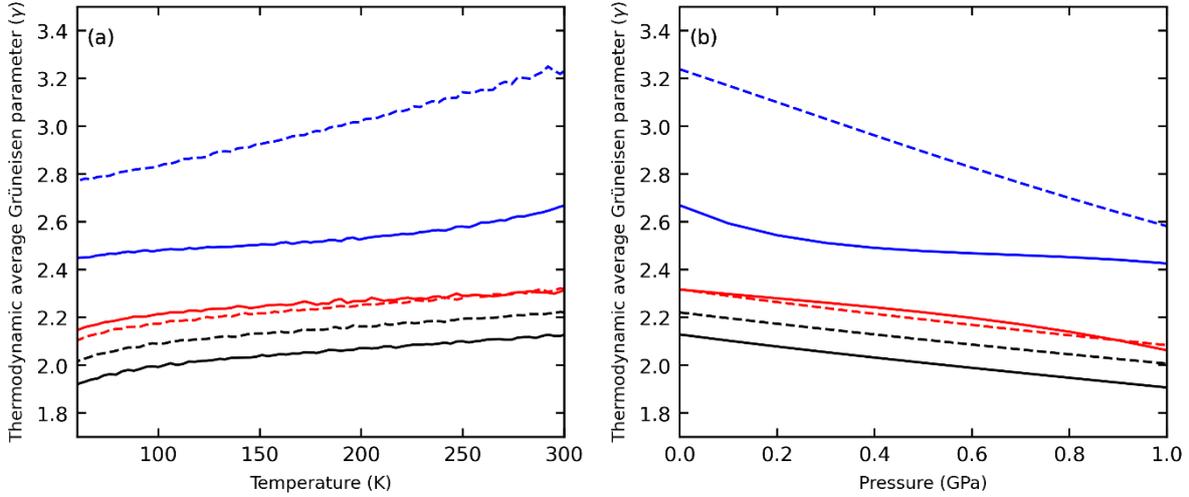

**Fig. 8:** (a) Temperature and (b) Pressure-dependent thermodynamic average Grüneisen parameter (γ) at 0 kbar and 301 K, respectively. The solid and dotted lines represent the fully relativistic and scalar relativistic cases, and the black, blue, and red represent the LDA, PBE, and PBEsol.

Table 3 presents the equilibrium elastic constant coefficients ($C_{ij}$) of polonium calculated at 0 K for PBE, PBEsol, and LDA, both with and without SOC. Using $C_{11}$ and $C_{12}$, we calculated the bulk modulus K = $(C_{11}+2C_{12})/3$ (shown in Table 3) and found that these are very close to the value presented in Table 1. Since SC polonium has a strong elastic anisotropy [10], we also calculated and compared the tetragonal shear moduli $C'=(C_{11}-C_{12})/2$ and elastic anisotropy factor A= $C_{44}/C'$. Here, A=1 represents elastic isotropic, and A≠1 indicates an anisotropic system (specifically, A<1 or A>1 implies lower or higher resistance to shear in {100} planes). In Table 3, our results show A<< 1, indicating a strong elastic anisotropic nature of simple cubic polonium, which is consistent with all-electron [10] and tight-binding [38] studies. Also, introducing the SOC, the $C_{ij}$, K, C′, and A decreases for all the functionals. With SOC, the trigonal shear modulus $C_{44}$ reduces by 42 % for LDA, 31% for PBE and 39 % for PBEsol. Additionally, the C′ decreases by 24 %, 19%, and 17% for PBE, PBEsol, and LDA, respectively, whereas the A decreases only by 9 % for PBE and 33 % for PBEsol and LDA, when SOC is included. These decreases in the values are consistent with the tight-binding (TB) results[38] and with the LDA and LDA+SOC results of all-electron calculations [10]. However, in Ref.[10], the value of $C_{44}$ increases for GGA when SOC is considered. To understand this discrepancy, we calculated the elastic constant coefficients at the same equilibrium lattice constant with PAW pseudopotentials; a similar SOC-induced decrease in $C_{ij}$ was observed (not shown here).

|  |  | Elastic constant coefficients ($C_{ij}$) in GPa | | | Bulk modulus (K) in GPa | Tetragonal shear moduli (C′) in GPa | Elastic anisotropy factor (A) |
|---|---|---|---|---|---|---|---|
|  |  | $C_{11}$ | $C_{12}$ | $C_{44}$ | K=($C_{11}$+2$C_{12}$)/3 | C′=($C_{11}$-$C_{12}$)/2 | A=$C_{44}$/ C′ |
| This study (US-PP) | LDA | 109.6 | 28.8 | 2.4 | 55.7 | 40.4 | 0.06 |
|  | LDA+SOC | 87.3 | 20.5 | 1.4 | 42.8 | 33.4 | 0.04 |
|  | PBE | 89.1 | 23.6 | 3.6 | 45.4 | 32.75 | 0.11 |
|  | PBE+SOC | 65.4 | 15.5 | 2.5 | 32.1 | 24.95 | 0.10 |
|  | PBEsol | 104.5 | 26.8 | 2.3 | 52.7 | 38.85 | 0.06 |
|  | PBEsol+SOC | 81.4 | 18.6 | 1.4 | 39.5 | 31.4 | 0.04 |
| Other Studies | LDA (FLAPW)[10] | - | - | 3.47 | - | 48.30 | 0.072 |
|  | LDA+SOC (FLAPW)[10] | - | - | 2.17 | - | 34.40 | 0.063 |
|  | LDA (NC-PP)[3] | 113 | 28 |  | 56 | 42.50 | - |
|  | GGA (FLAPW)[10] | - | - | 3.54 | - | 36.60 | 0.097 |
|  | GGA+SOC (FLAPW)[10] | - | - | 4.55 | - | 26.80 | 0.170 |
|  | HL (TB) [38] | 138 | 20 | 2 | 59.3 | 59.0 | 0.034 |
|  | HL+SOC (TB) [38] | 121 | 16 | 2 | 51.0 | 52.5 | 0.038 |
|  | HL (LAPW) [38] | 92 | 20 | 5 | 44.0 | 36.0 | 0.139 |

**Table 3:** The equilibrium elastic constant coefficients ($C_{ij}$) for LDA, PBE, and PBEsol with and without SOC. The bulk modulus K = ($C_{11}$+2$C_{12}$)/3, tetragonal shear moduli C′=($C_{11}$-$C_{12}$)/2, and elastic anisotropy factor A=$C_{44}$/ C′, where $C_{44}$ is the trigonal shear modulus, are calculated. The available theoretical results from Ref.[10], Ref.[38], and Ref.[3] are compared.

Fig.9 shows the pressure-dependent elastic constant coefficient ($C_{ij}$) and elastic anisotropic factor (A) at 0 K, calculated on the geometries used for QHA calculations for LDA and LDA+SOC. We also included the $C_{ij}$ at equilibrium conditions. Corresponding results for PBE and PBEsol are included in supplementary S8. In all cases, the $C_{ij}$ satisfies the Born stability criteria [44] for a simple cubic system ($C_{11}$ > 0, $C_{11}$–$C_{12}$ > 0, $C_{11}$ + 2$C_{12}$ > 0, and $C_{44}$ > 0). Together with the positive phonon frequencies, the agreement with stability criteria further assures that all the studied geometries used to study the thermodynamic properties using different functionals, both with and without SOC, are stable. As described in Fig.9(a), $C_{11}$ and $C_{12}$ increase with the pressure, while $C_{44}$ decreases for both with and without SOC cases and all functionals (see supplementary S8). Consequently, the decrease in the elastic anisotropic factor indicates that with increasing pressure (or decreasing the atomic volume), the simple cubic structure loses its stability due to trigonal deformation[10].

Finally, we investigate the pressure-dependent behaviour of Cauchy pressure ($P_C$) and Pugh ratio. The Cauchy pressure, defined as $P_C = C_{12} - C_{44}$, indicates the degree of angularity or directionality in the bonding of crystalline solids. Positive $P_C$ refers to a delocalized, isotropic electronic cloud (metallic bonding), and negative $P_C$ indicates directional bonding in covalent or ionic solids. Fig.10(a) shows that our equilibrium LDA+SOC is higher than the all-electron[11] result. Also, the $P_C$ is highly positive for all cases, referring to the strong metallic character of the polonium. The value of $P_C$ for all the functionals decreases when SOC is included. Notably, we found that the difference in the $P_C$ obtained with and without SOC, at equilibrium (0 GPa), is ~ 7 GPa irrespective of the functional used. This indicates that SOC introduces a uniform weakening of shear resistance ($C_{44}$) relative to lateral bonding stiffness ($C_{12}$) without altering the overall bonding nature, i.e., metallic. To summarize, the marginal difference in bulk modulus, and a significant change in the tetragonal and trigonal shear modulus upon the inclusion of SOC. This indicates the SOC primarily affects directional bonding rather than isotropic compression

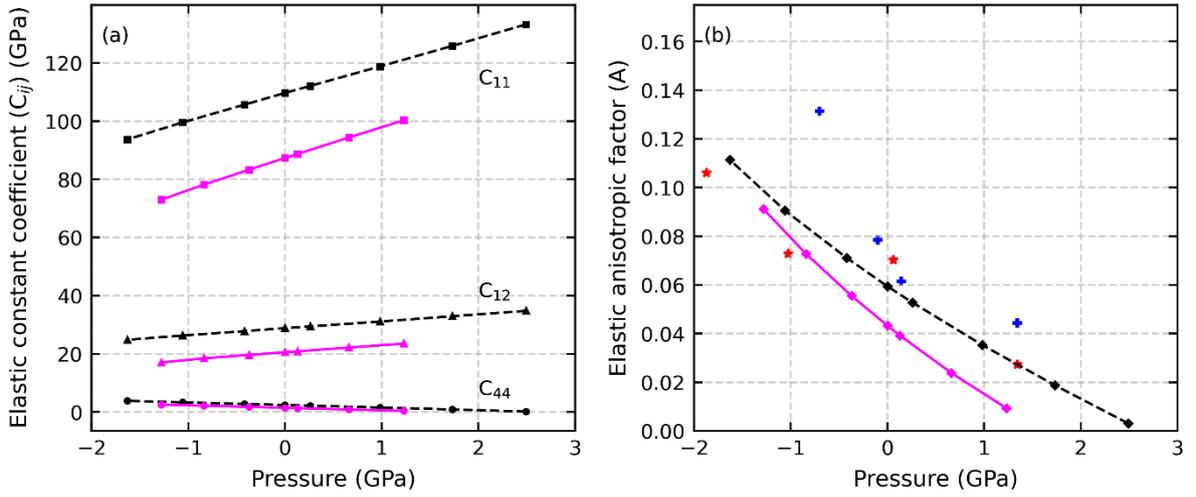

**Fig.9:** (a) Elastic constant coefficient ($C_{ij}$) and (b) elastic anisotropic factor (A) as a function of pressure for LDA (black broken lines) and LDA+SOC (magenta solid lines). The $C_{11}$ (square), $C_{12}$ (triangle), $C_{44}$ (circle), and A are calculated on the equilibrium (0 GPa) geometry, and the other geometries are used for studying thermodynamics using QHA. In (b), the symbols ∗ and + are LDA and LDA+SOC results from Ref.[10].

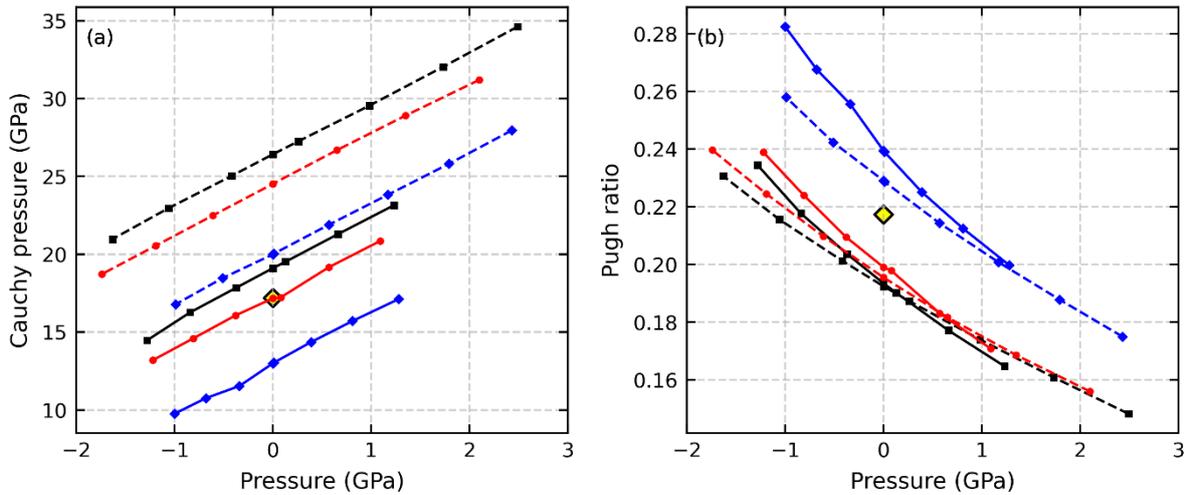

**Fig.10:** Pressure-dependent (a) Cauchy pressure ($P_C = C_{12} - C_{44}$) and (b) Pugh ratio (G/B) for LDA (black square), PBEsol (red circle), and PBE (blue diamond) determined using (solid lines) and without (dashed) SOC. The value from the LDA+SOC result from the all-electron method [11] is shown in a yellow diamond.

Fig.10(b) presents the Pugh ratio for PBE, PBEsol, and LDA at 0 K, both with and without SOC. The Pugh ratio (G/B) is obtained from the polycrystalline bulk modulus (B) and shear modulus (G) using the single crystals $C_{ij}$'s and Voigt-Reuss-Hill (VRH) averaging relations [36]. At equilibrium, the Pugh ratios are ~0.19 for LDA, ~0.24 for PBE, and ~0.20 for PBEsol, whereas without SOC, they are 0.19, ~ 0.23, and ~0.20, respectively. These ratios are lower than the critical Pugh ratio of 0.57 (< for ductile and > for brittle), indicating that the polonium at equilibrium conditions is ductile. This result is consistent with the Legut *et al.* [11] where the reported Pugh ratio for LDA+SOC is 0.22. Moreover, the incorporation of SOC reduces the ductility. As the pressure increases, the Pugh ratio decreases for all cases, indicating enhanced ductility, as illustrated in Fig.9(b).

**Conclusions:**

This work presents a comprehensive first-principles study of the role of SOC on the thermodynamic and mechanical properties of simple cubic polonium by comparing the results from scalar-relativistic (no-SOC) and fully relativistic (with-SOC) pseudopotentials. Additionally, we examined the performance of different exchange-correlation functionals (LDA, PBE, and PBEsol) with both PPs. The contribution of both lattice vibration and electronic excitations is included in the analysis, and we found that the latter contribution is negligible.

Our results indicate that LDA+SOC accurately reproduces the experimental volume, whereas both LDA+SOC and PBEsol + SOC successfully give thermal expansion coefficients very close to the experimental and other theoretical reports. A systematic difference in β between results with and without SOC was observed across the entire temperature range, and this difference is independent of the functional. At 301 K, the contribution of SOC in β increases with pressure for PBE and PBEsol, while for LDA, it decreases. In contrast, the SOC effects are insignificant in isobaric heat capacity, with LDA and PBEsol both with and without SOC agreeing with the experimental values. Also, with pressure, the influence of SOC on $C_P$ is negligible (~ 1%) for all the functionals. In contrast, SOC significantly affects the adiabatic bulk modulus, resulting in reductions of over 20% depending on the functional. The phonon dispersions illustrate several anomalies in all directions, and they are reduced significantly when SOC is considered. However, we found that, partly due to variations in lattice constants used, the SOC effect on phonon frequencies differs between different studies. When computed at the experimental lattice constant, the present results agree with Ref.[6], confirming that SOC mainly impacts transverse phonons while having minimal influence on longitudinal modes, except at **R**.

Mechanical properties analysis at 0 K indicates that SOC lowers the value of elastic constant coefficients ($C_{ij}$), trigonal shear modulus (C′), elastic anisotropic factor (A), Cauchy pressure ($P_C$), and Pugh ratio. The elastic anisotropic factor decreases with increasing pressure, indicating the simple cubic structure loses its stability due to trigonal deformation. The positive $P_C$ for all cases affirms the strong metallic character of the polonium. The existence of a uniform difference of approximately 7 GPa in the $P_C$ obtained with and without SOC for all the functionals implies that SOC uniformly weakens shear resistance ($C_{44}$) without changing the overall bonding nature. Moreover, $C_{11}$ and $C_{12}$ increase with increasing pressure while $C_{44}$ decreases, reducing the anisotropy factor A, indicating a tendency towards trigonal instability at high pressure. The Pugh ratio indicates that polonium is ductile at equilibrium pressure, and SOC slightly reduces the ductility. Furthermore, with increasing pressure, there prevails an enhancement in the ductility at 0 K. To conclude, our findings highlight the significant role of SOC on polonium's vibrational and mechanical properties and provide a theoretical platform for understanding its stability and bonding nature under varying temperature and pressure conditions.

**Conflicts of interest:**

There are no conflicts to declare.

**Data availability:**

All data supporting the findings of this study are included in the main article and the Supplementary Information.

**Acknowledgments:**

This work has been supported from the European Union through the MAX "MAterials design at the eXascale" Centre of Excellence for Supercomputing applications (Grant agreement No. 101093374, co-funded by the European High Performance Computing joint Undertaking (JU) and participating countries 824143). Computational facilities have been provided by SISSA through its Linux Cluster, ITCS, and the SISSA-CINECA 2021-2025 Agreement. Partial support has been received by the Italian MUR (Ministry of University and Research) through the National Centre for HPC, Big Data, and Quantum Computing (grant No. CN00000013).

**References:**

1   M. I. McMahon and R. J. Nelmes, High-pressure structures and phase transformations in elemental metals, *Chem. Soc. Rev.*, 2006, **35**, 943.

2   David R. Lide, Ed., *CRC Handbook of Chemistry and Physics*, CRC Press Inc, Boca Raton, Florida, 69th edn., 2005.

3   R. E. Kraig, D. Roundy and M. L. Cohen, A study of the mechanical and structural properties of polonium, *Solid State Commun.*, 2004, **129**, 411–413.

4   A. Zaoui, A. Belabbes, R. Ahuja and M. Ferhat, Interplay between lattice dynamics and the low-pressure phase of simple cubic polonium, *Phys. Lett. Sect. A Gen. At. Solid State Phys.*, 2011, **375**, 1695–1697.

5   C.-J. Kang, K. Kim and B. I. Min, Phonon softening and superconductivity triggered by spin-orbit coupling in simple-cubic α-polonium crystals, *Phys. Rev. B*, 2012, **86**, 054115.

6   M. J. Verstraete, Phases of Polonium via Density Functional Theory, *Phys. Rev. Lett.*, 2010, **104**, 035501.

7   A. Belabbes, A. Zaoui and M. Ferhat, Strong phonon anomalies and Fermi surface nesting of simple cubic Polonium, *Solid State Commun.*, 2010, **150**, 2337–2340.

8   A. Dal Corso, Ab initio phonon dispersions of face centered cubic Pb: effects of spin–orbit coupling, *J. Phys. Condens. Matter*, 2008, **20**, 445202.

9   B. Thakur, X. Gong and A. Dal Corso, Thermodynamic properties of fcc lead: A scalar and fully relativistic first principle study, *Comput. Mater. Sci.*, 2025, **249**, 113677.

10  D. Legut, M. Friák and M. Šob, Why Is Polonium Simple Cubic and So Highly Anisotropic?, *Phys. Rev. Lett.*, 2007, **99**, 016402.

11  D. Legut, M. Friák and M. Šob, Phase stability, elasticity, and theoretical strength of polonium from first principles, *Phys. Rev. B - Condens. Matter Mater. Phys.*, 2010, **81**, 214118.

12  A. Dal Corso, The thermo_pw code can be downloaded from the web page, https://dalcorso.github.io/thermo_pw/.

13  P. Giannozzi, S. Baroni, N. Bonini, M. Calandra, R. Car, C. Cavazzoni, D. Ceresoli, G. L. Chiarotti, M. Cococcioni, I. Dabo, A. Dal Corso, S. de Gironcoli, S. Fabris, G. Fratesi, R. Gebauer, U. Gerstmann, C. Gougoussis, A. Kokalj, M. Lazzeri, L. Martin-Samos, N. Marzari, F. Mauri, R. Mazzarello, S. Paolini, A. Pasquarello, L. Paulatto, C. Sbraccia, S. Scandolo, G. Sclauzero, A. P. Seitsonen, A. Smogunov, P. Umari, R. M. Wentzcovitch and P. U. and R. M. W. Paolo Giannozzi, Stefano Baroni, Nicola Bonini, Matteo Calandra, Roberto Car, Carlo Cavazzoni, Davide Ceresoli, Guido L Chiarotti, Matteo Cococcioni, Ismaila Dabo, Andrea Dal Corso, Stefano de Gironcoli, Stefano Fabris, Guido Fratesi, Ralph Gebauer, Uwe G,

QUANTUM ESPRESSO: a modular and open-source software project for quantum simulations of materials, *J. Phys. Condens. Matter*, 2009, **21**, 395502.

14  P. Giannozzi, O. Andreussi, T. Brumme, O. Bunau, M. Buongiorno Nardelli, M. Calandra, R. Car, C. Cavazzoni, D. Ceresoli, M. Cococcioni, N. Colonna, I. Carnimeo, A. Dal Corso, S. de Gironcoli, P. Delugas, R. A. DiStasio, A. Ferretti, A. Floris, G. Fratesi, G. Fugallo, R. Gebauer, U. Gerstmann, F. Giustino, T. Gorni, J. Jia, M. Kawamura, H.-Y. Ko, A. Kokalj, E. Küçükbenli, M. Lazzeri, M. Marsili, N. Marzari, F. Mauri, N. L. Nguyen, H.-V. Nguyen, A. Otero-de-la-Roza, L. Paulatto, S. Poncé, D. Rocca, R. Sabatini, B. Santra, M. Schlipf, A. P. Seitsonen, A. Smogunov, I. Timrov, T. Thonhauser, P. Umari, N. Vast, X. Wu and S. Baroni, Advanced capabilities for materials modelling with Quantum ESPRESSO, *J. Phys. Condens. Matter*, 2017, **29**, 465901.

15  D. Vanderbilt, Soft self-consistent pseudopotentials in a generalized eigenvalue formalism, *Phys. Rev. B*, 1990, **41**, 7892–7895.

16  A. Dal Corso, The pslibrary pseudopotential library, https://github.com/dalcorso/pslibrary/.

17  A. Dal Corso, Pseudopotentials periodic table: From H to Pu, *Comput. Mater. Sci.*, 2014, **95**, 337–350.

18  A. M. Rappe, K. M. Rabe, E. Kaxiras and J. D. Joannopoulos, Optimized pseudopotentials, *Phys. Rev. B*, 1990, **41**, 1227–1230.

19  A. Dal Corso and A. M. Conte, Spin-orbit coupling with ultrasoft pseudopotentials: Application to Au and Pt, *Phys. Rev. B*, 2005, **71**, 115106.

20  A. Dal Corso, Density functional perturbation theory for lattice dynamics with fully relativistic ultrasoft pseudopotentials: Application to fcc-Pt and fcc-Au, *Phys. Rev. B*, 2007, **76**, 054308.

21  J. P. Perdew and A. Zunger, Self-interaction correction to density-functional approximations for many-electron systems, *Phys. Rev. B*, 1981, **23**, 5048–5079.

22  J. P. Perdew, K. Burke and M. Ernzerhof, Generalized Gradient Approximation Made Simple, *Phys. Rev. Lett.*, 1996, **77**, 3865–3868.

23  J. P. Perdew, A. Ruzsinszky, G. I. Csonka, O. A. Vydrov, G. E. Scuseria, L. A. Constantin, X. Zhou and K. Burke, Restoring the Density-Gradient Expansion for Exchange in Solids and Surfaces, *Phys. Rev. Lett.*, 2008, **100**, 136406.

24  S. G. Louie, S. Froyen and M. L. Cohen, Nonlinear ionic pseudopotentials in spin-density-functional calculations, *Phys. Rev. B*, 1982, **26**, 1738–1742.

25  H. J. Monkhorst and J. D. Pack, Special points for Brillouin-zone integrations, *Phys. Rev. B*, 1976, **13**, 5188–5192.

26  M. Methfessel and A. T. Paxton, High-precision sampling for Brillouin-zone integration in metals, *Phys. Rev. B*, 1989, **40**, 3616–3621.

27  S. Baroni, S. de Gironcoli, A. Dal Corso and P. Giannozzi, Phonons and related crystal properties from density-functional perturbation theory, *Rev. Mod. Phys.*, 2001, **73**, 515–562.

28  A. Dal Corso, Density-functional perturbation theory with ultrasoft pseudopotentials, *Phys. Rev. B*, 2001, **64**, 235118.

29  C. Malica and A. Dal Corso, Temperature dependent elastic constants and thermodynamic properties of BAs: An ab initio investigation, *J. Appl. Phys.*, 2020, **127**, 245103.

30  C. Malica and A. Dal Corso, Quasi-harmonic thermoelasticity of palladium, platinum, copper, and gold from first principles, *J. Phys. Condens. Matter*, 2021, **33**, 475901.

31  A. Dal Corso, Elastic constants of beryllium: a first-principles investigation, *J. Phys. Condens. Matter*, 2016, **28**, 075401.

32  B. Thakur, X. Gong and A. Dal Corso, Ab initio thermodynamic properties of iridium: A high-pressure and high-temperature study, *Comput. Mater. Sci.*, 2024, **234**, 112797.

33 B. Thakur, X. Gong and A. Dal Corso, Thermodynamic properties of rhodium—A first principle study, *AIP Adv.*, 2024, **14**, 045229–9.

34 X. Gong and A. Dal Corso, Ab initio quasi-harmonic thermoelasticity of molybdenum at high temperature and pressure, *J. Chem. Phys.*, 2024, **160**, 244703–6.

35 X. Gong and A. Dal Corso, Pressure and temperature dependent ab-initio quasi-harmonic thermoelastic properties of tungsten, *J. Phys. Condens. Matter*, 2024, **36**, 285702.

36 R. Hill, The Elastic Behaviour of a Crystalline Aggregate, *Proc. Phys. Soc. Sect. A*, 1952, **65**, 349–354.

37 W. H. Beamer and C. R. Maxwell, The Crystal Structure of Polonium, *J. Chem. Phys.*, 1946, **14**, 569–569.

38 M. Lach-hab, B. Akdim, D. A. Papaconstantopoulos, M. J. Mehl and N. Bernstein, Application of the NRL tight-binding method to the heavy elements Pb and Po, *J. Phys. Chem. Solids*, 2004, **65**, 1837–1842.

39 R. J. DeSando and R. C. Lange, The structures of polonium and its compounds—I α and β polonium metal, *J. Inorg. Nucl. Chem.*, 1966, **28**, 1837–1846.

40 R. E. Brocklehurst, J. M. Goode and L. F. Vassamillet, Coefficient of Expansion of Polonium, *J. Chem. Phys.*, 1957, **27**, 985–985.

41 J. A. Dean, *Lange's handbook of chemistry*, McGRAW-HILL, INC., New York, 15th edn., 1999.

42 B. I. Min, J. H. Shim, M. S. Park, K. Kim, S. K. Kwon and S. J. Youn, Origin of the stabilized simple-cubic structure in polonium: Spin-orbit interaction versus Peierls instability, *Phys. Rev. B*, 2006, **73**, 132102.

43 N. W. Ashcroft and N. D. Mermin, *Solid State Physics*, Saunders College Publishing, New York, College Ed., 1976.

44 G. Grimvall, B. Magyari-Köpe, V. Ozoliņš and K. A. Persson, Lattice instabilities in metallic elements, *Rev. Mod. Phys.*, 2012, **84**, 945–986.

Supplementary data

# Scalar and fully relativistic pressure and temperature-dependent *ab-initio* thermodynamics study of simple cubic polonium

Balaram Thakur[1*], Xuejun Gong[1,2], and Andrea Dal Corso[1,3]

[1]International School for Advanced Studies (SISSA), Via Bonomea 265, 34136 Trieste, Italy.

[2]School of Physical Science and Technology, Xinjiang University, Urumqi, Xinjiang, 830046, China.

[3]CNR-IOM, Via Bonomea 265, 34136 Trieste, Italy.

*Corresponding author: bmtr789@gmail.com

Email:   Balaram Thakur (bmtr789@gmail.com), Xuejun Gong (xgong@sissa.it), Andrea Dal Corso (dalcorso@sissa.it)

**S1: Convergence Tests:**

The convergence tests were performed for the LDA, PBEsol, and PBE functionals, both with and without spin-orbit coupling (SOC). The convergence test was performed at 6.32185 a.u. The pseudo-wavefunctions and charge densities were expanded in a plane wave basis set with a kinetic energy cutoff of 80 Ry and 400 Ry, respectively. The convergence test for LDA(PZ), PBEsol, and PBE is shown in Fig.S1(A). With these energy cutoffs, the **k**-mesh and smearing factors convergence tests were performed and are shown in Figs. S1(B) to S1(J).

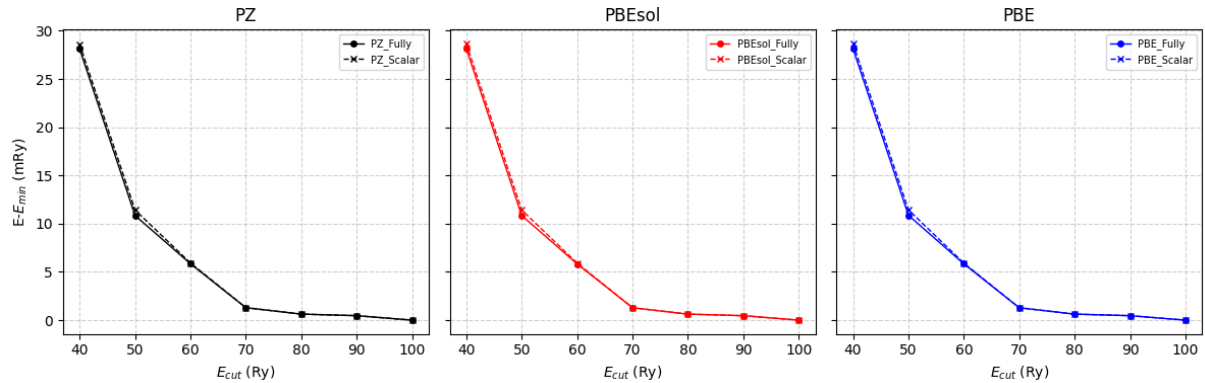

**Fig.S1(A):** Pseudo-wavefunction kinetic energy cutoff convergence test for LDA, PBEsol and PBE with fully (with SOC) and scalar (without SOC) pseudopotentials.

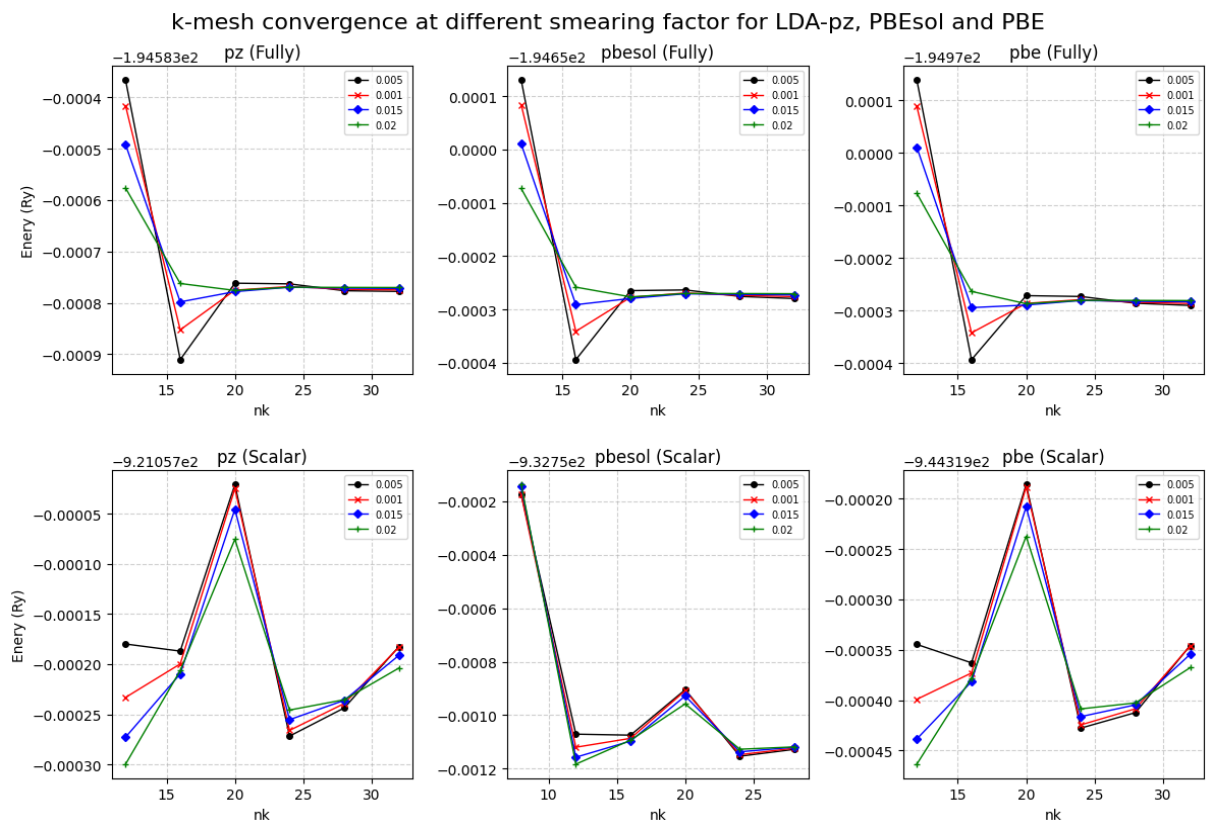

**Fig.S1(B):** k-mesh convergence tests for LDA, PBEsol, and PBE with fully (with SOC) and scalar (without SOC) pseudopotentials, at different MP smearing factors.

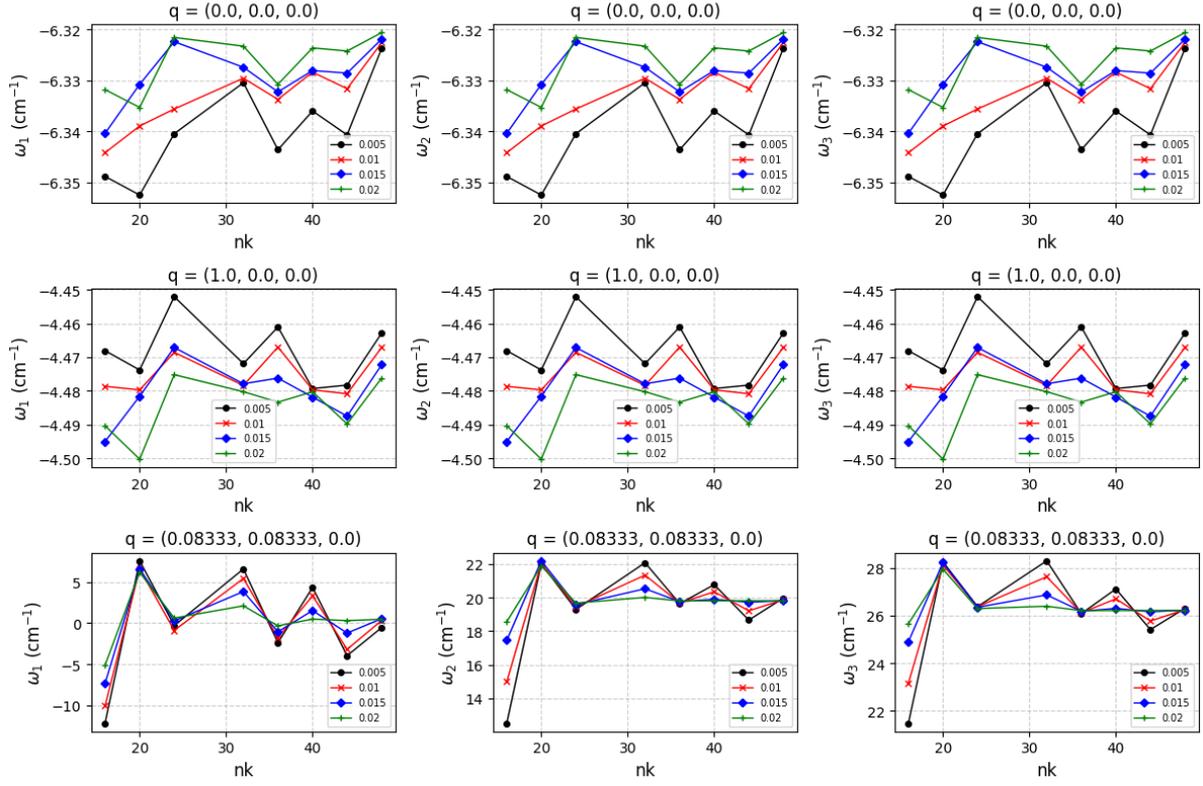

**Fig. S1(C):** Phonon frequency convergence tests at **q** (0,0,0), (1,0,0) and (1/12, 1/12, 0) or (0.08333, 0.08333, 0) at different MP smearing factor for LDA with SOC.

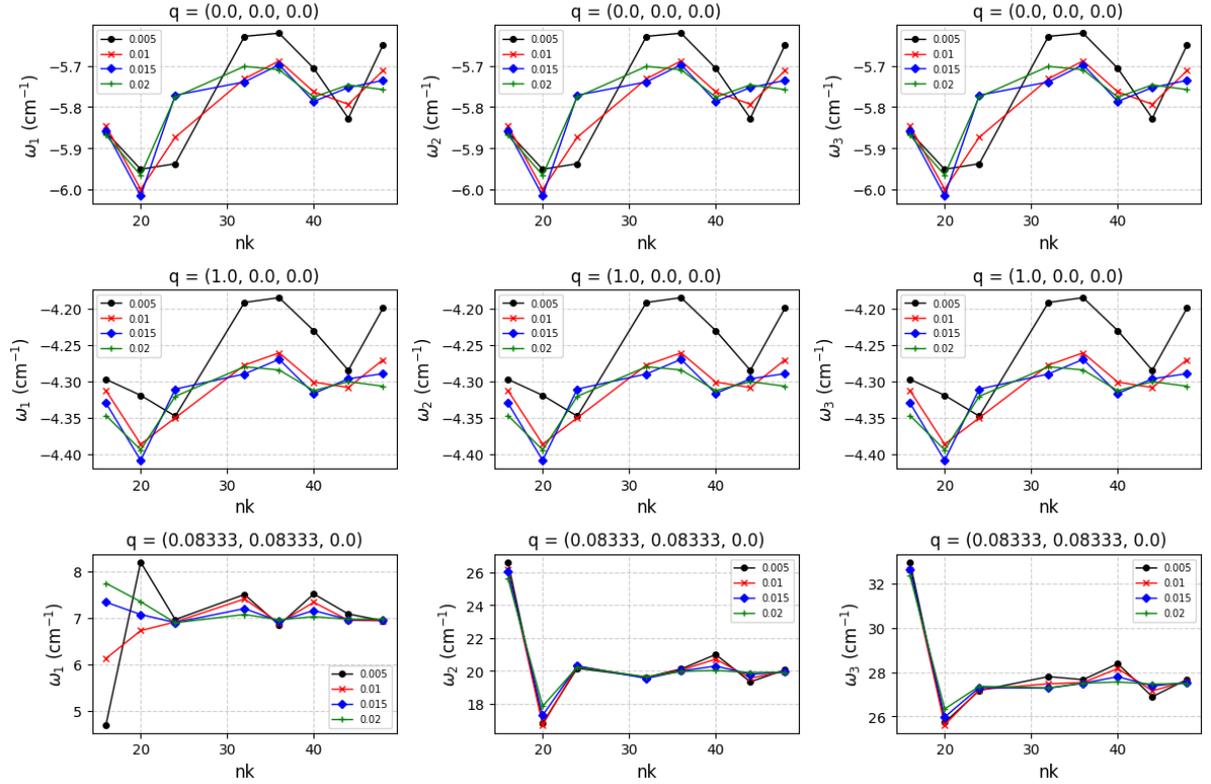

**Fig. S1(D):** Phonon frequency convergence tests at **q** (0,0,0), (1,0,0) and (1/12, 1/12, 0) or (0.08333, 0.08333, 0) at different MP smearing factor for LDA without SOC.

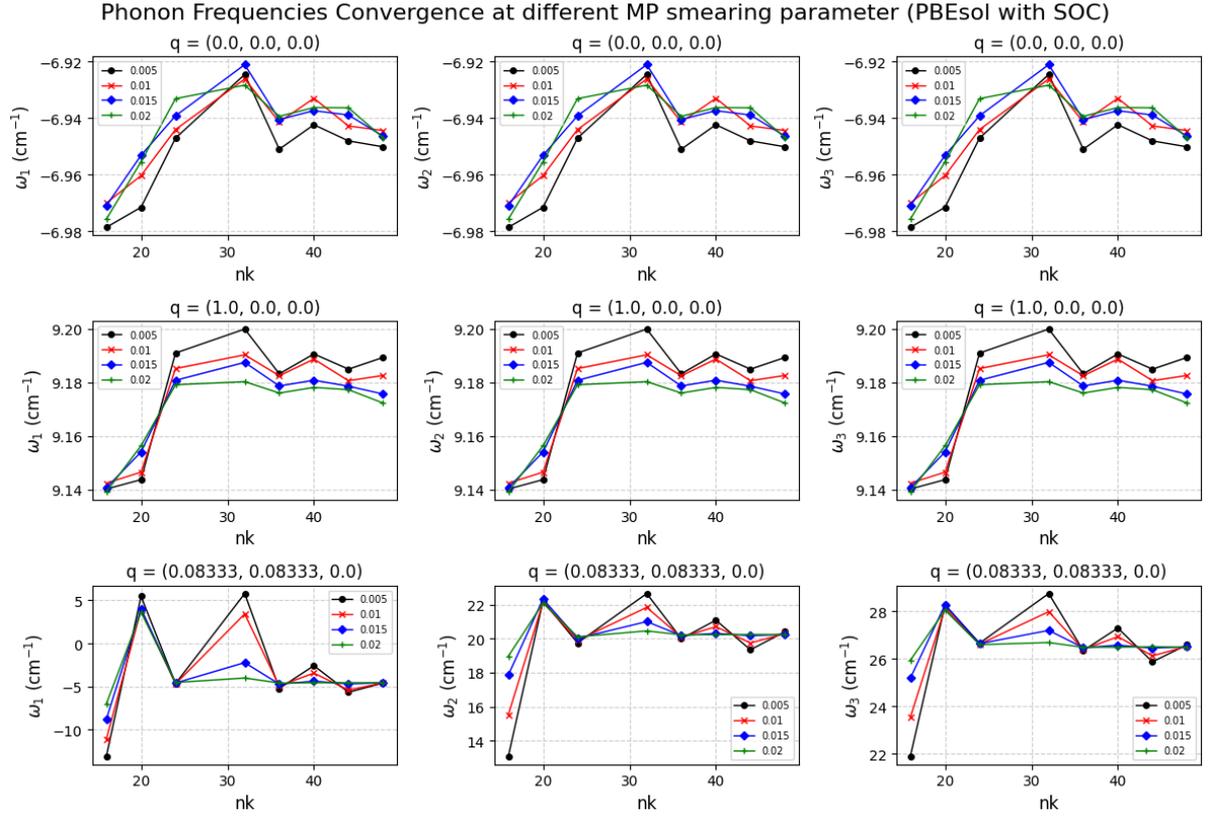

**Fig. S1(E):** Phonon frequency convergence tests at **q** (0,0,0), (1,0,0) and (1/12, 1/12, 0) or (0.08333, 0.08333, 0) at different MP smearing factors for PBEsol with SOC.

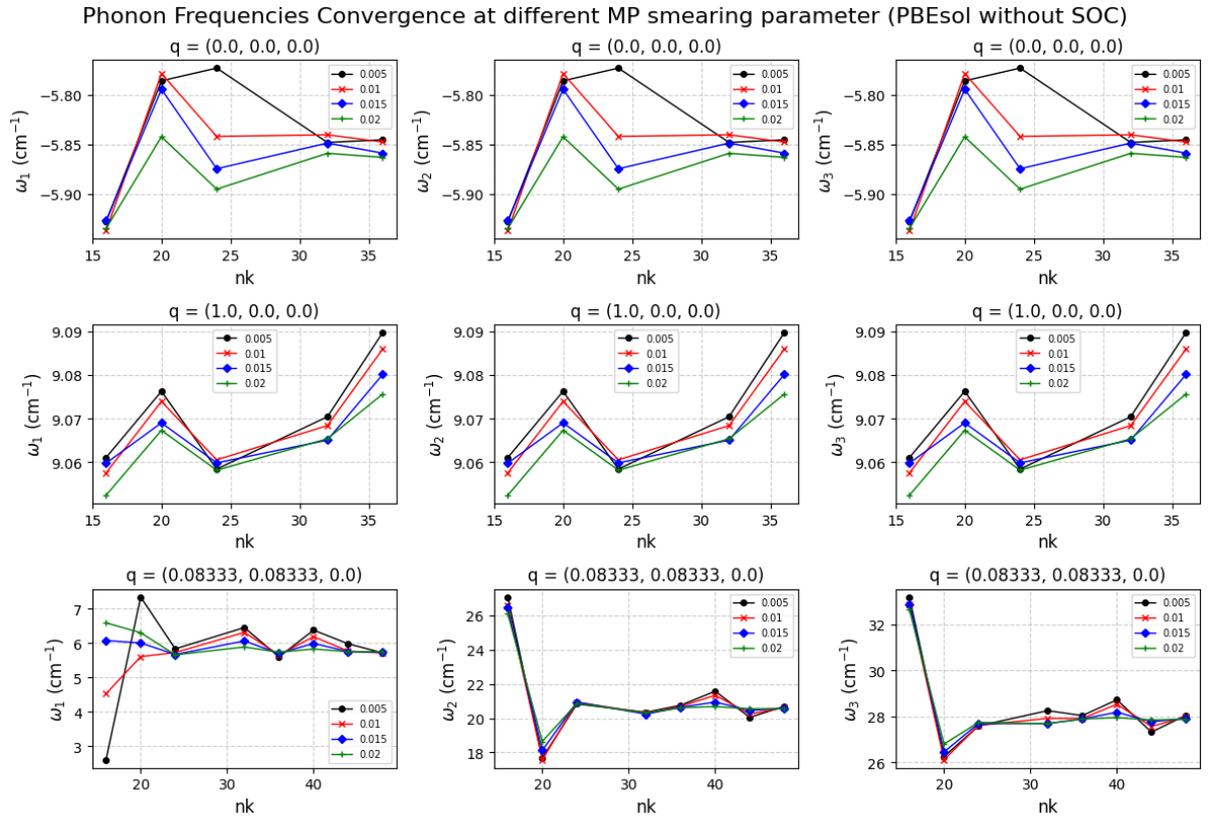

**Fig. S1(F):** Phonon frequency convergence tests at **q** (0,0,0), (1,0,0) and (1/12, 1/12, 0) or (0.08333, 0.08333, 0) at different MP smearing factors for PBEsol without SOC.

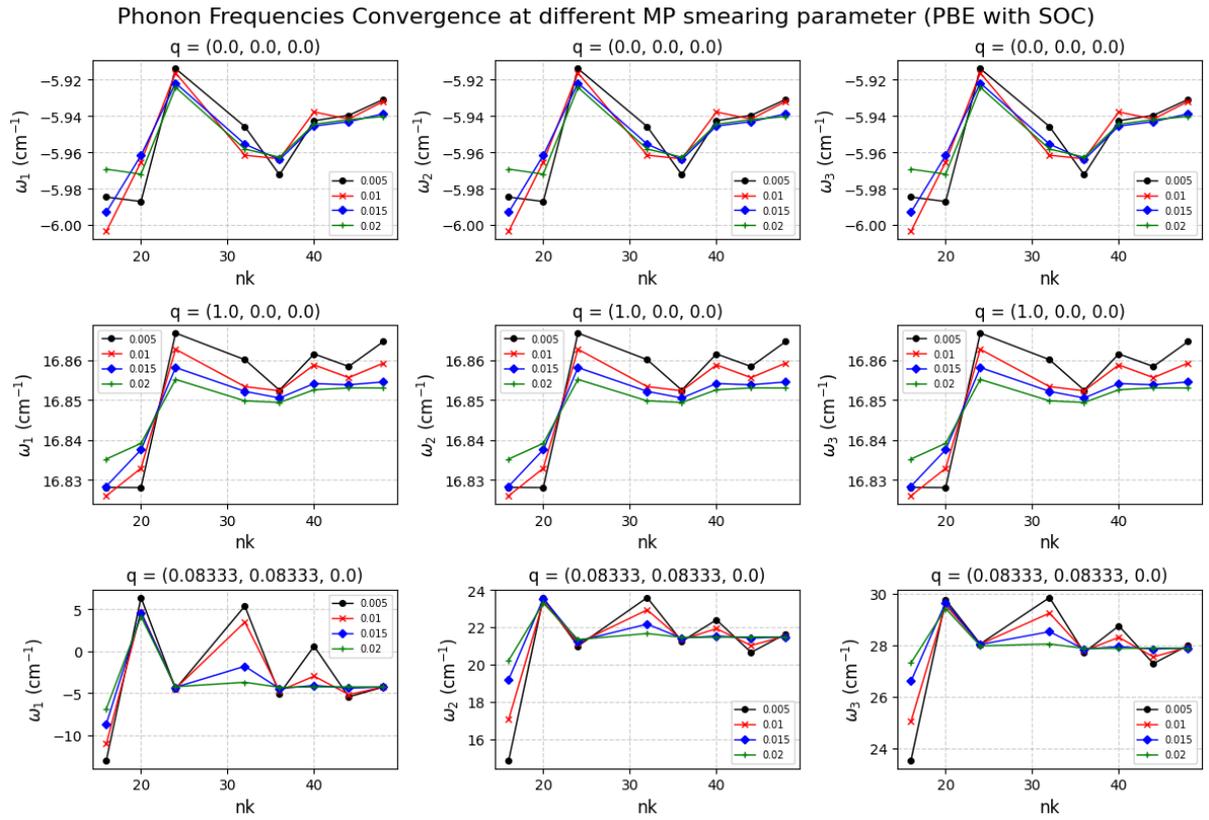

**Fig. S1(G):** Phonon frequency convergence tests at **q** (0,0,0), (1,0,0) and (1/12, 1/12, 0) or (0.08333, 0.08333, 0) at different MP smearing factors for PBE with SOC.

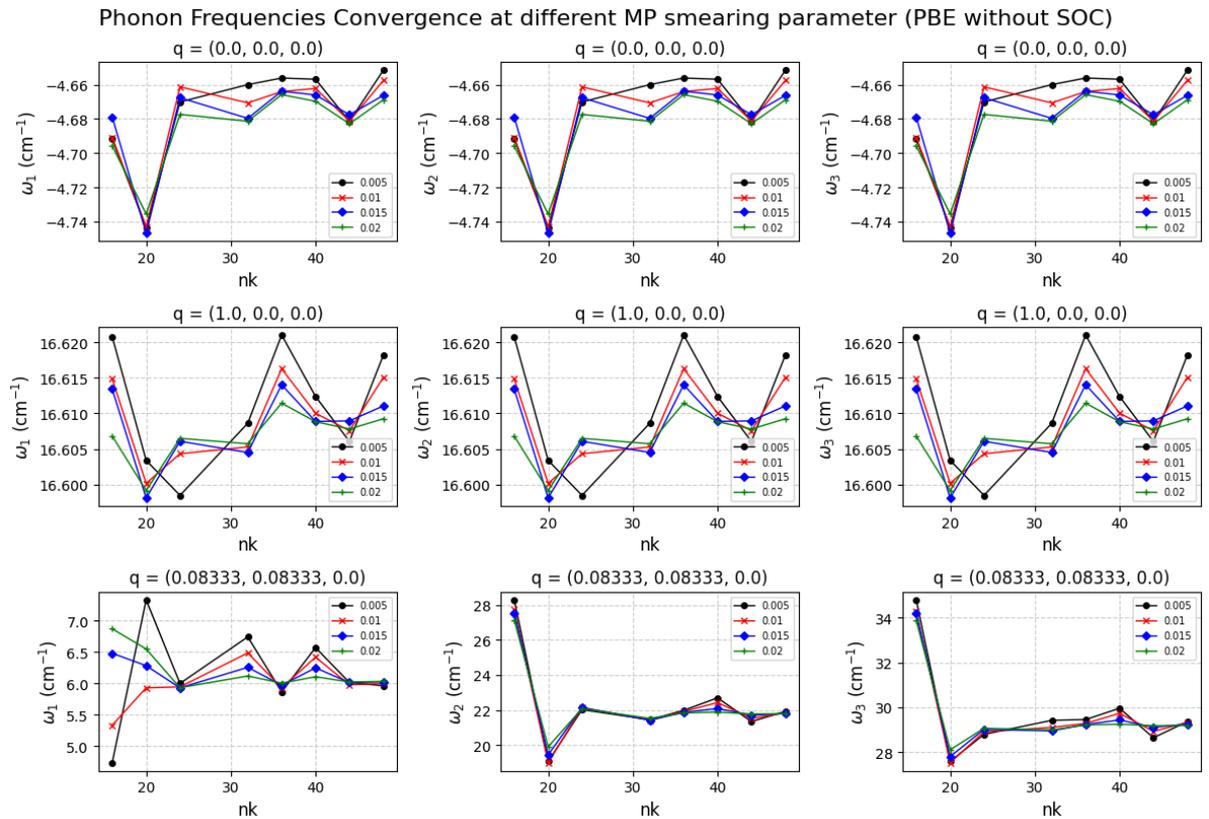

**Fig. S1(H):** Phonon frequency convergence tests at **q** (0,0,0), (1,0,0) and (1/12, 1/12, 0) or (0.08333, 0.08333, 0) at different MP smearing factors for PBE without SOC.

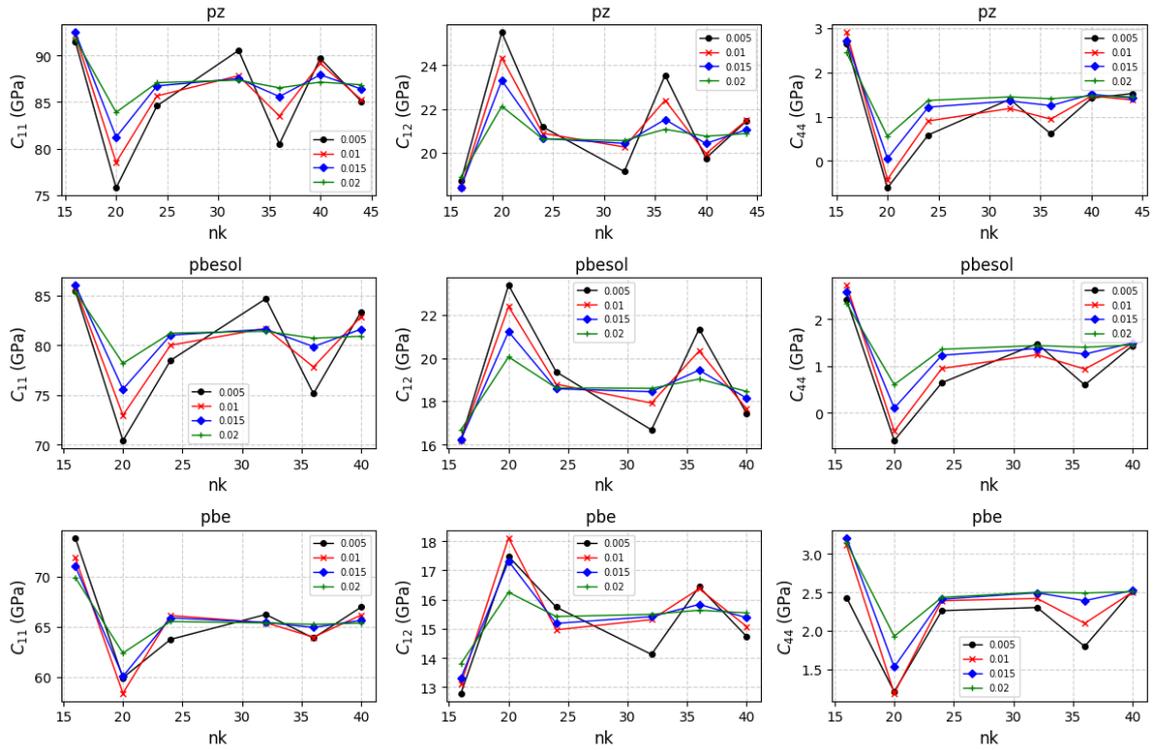

**Fig. S1(I):** Elastic constants coefficient convergence tests at different MP smearing factors for LDA, PBEsol, and PBE with SOC.

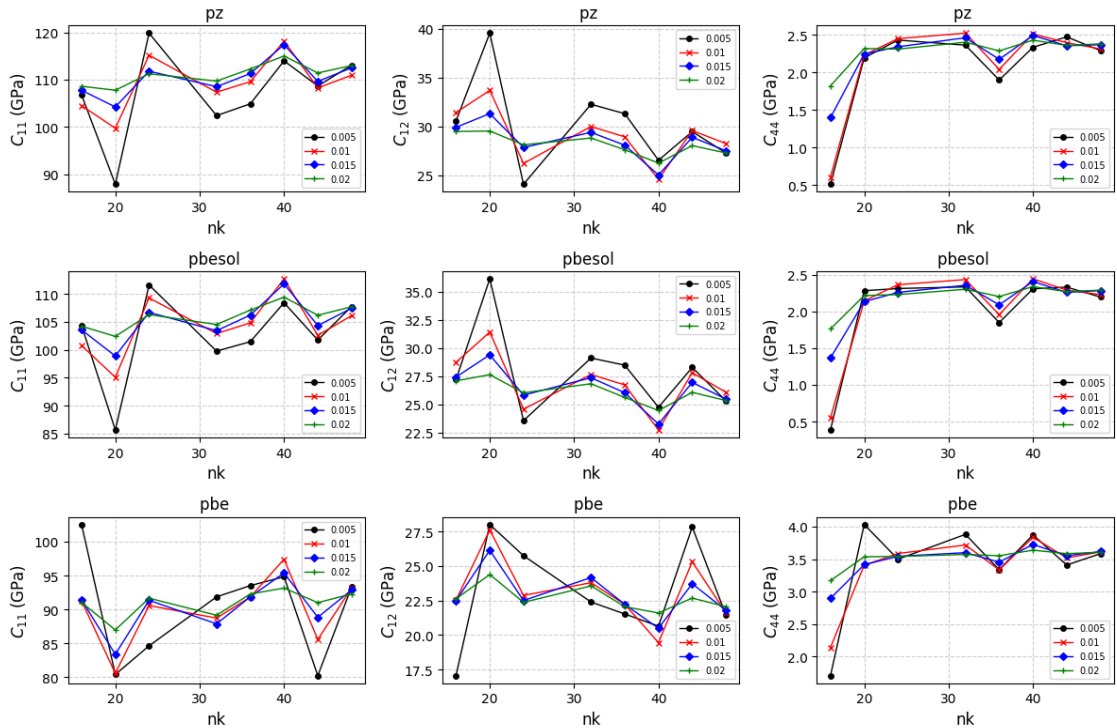

**Fig.S1(J):** Elastic constants coefficients convergence tests at different MP smearing factors for LDA, PBEsol, and PBE without SOC.

**S2: Details of the geometries on which dynamical matrices are calculated to determine the thermodynamic properties within quasi-harmonic approximations.**

| Methods | | No. of geometries | Step size (in a.u.) | Lattice constant range (in a.u.) |
|---|---|---|---|---|
| Scalar relativistic (without spin-orbit coupling) | LDA | 7 | 0.025 | 6.105 to 6.255 |
| | PBE | 7 | 0.025 | 6.230 to 6.380 |
| | PBEsol | 7 | 0.025 | 6.155 to 6.305 |
| Fully relativistic (with spin-orbit coupling) | LDA | 6 | 0.025 | 6.230 to 6.355 |
| | PBE | 7 | 0.025 | 6.385 to 6.535 |
| | PBEsol | 6 | 0.025 | 6.280 to 6.405 |

**S3. Theory:**

At a fixed crystal volume $V$, the dynamical matrices are calculated on a coarse grid of phonon wavevectors $q$ using DFPT [1]. These matrices are first Fourier interpolated on a dense mesh and then diagonalized to obtain the eigenvectors and phonon frequencies $\omega_\eta(q, V)$, where η is the phonon-mode or branch index. In QHA, the phonon frequencies are calculated at different volumes, and this volume-dependent phonon frequencies $\omega_\eta(q, V_i)$ allows the inclusion of anharmonic effects indirectly through the volume dependence of vibrational properties. Finally, the Helmholtz free energy F (V, T) is determined using direct summation over phonon frequencies and is read as:

$$F(V,T) = U_0(V) + F_{ph}(V,T) + F_{elec}(V,T) \quad (1)$$

In equation (1), $U_0(V)$ is the static (0 K) total energy of a system obtained from DFT and is fitted to a 4$^{th}$-order Birch–Murnaghan equation of state to ensure a smooth interpolation in volume. $F_{ph}(V,T)$ represents the lattice vibrations (phonons) contribution to the free energy and is given by:

$$F_{ph}(V,T) = \tfrac{1}{2}\sum_{q,\eta} \hbar\omega_\eta(q,V) + k_B T \sum_{q,\eta} \ln\left[1 - e^{-\frac{\hbar\omega_\eta(q,V)}{k_B T}}\right] \quad (2)$$

In equation (2), the first term is zero-point energy, and the second corresponds to the vibrational free energy at finite temperatures. The phonon free energy is interpolated using a 4$^{th}$-degree polynomial of volume. Furthermore, in (1), the $F_{elec}(V,T)$ is the electronic contribution to the free energy and is given by:

$$F_{ele}(V,T) = U_{ele}(V,T) - T \cdot S_{ele}(V,T) \quad (3)$$

Where $U_{elec}(V,T)$ and $S_{elec}(V,T)$ are determined using the electronic density of states (el-DOS) obtained from DFT for each volume V, using the rigid-bands approximation. The rigid-bands approximation assumes that el-DOS does not vary significantly with temperature. This term is also interpolated in the 4$^{th}$-order polynomial of volume. After determining the total Helmholtz free energy F(V, T), the pressure is derived from the volume derivative of F (V, T), and the relation is given as $P = -\frac{dF(V,T)}{dV}$.

At each temperature and pressure, the equilibrium volume $V_{eq}(P,T)$ is obtained by minimizing the Gibbs free energy $G_P(V,T) = F(V,T) + P \cdot V$, where pressure P enters parametrically in the relation. Using the equilibrium volume $V_{eq}(P,T)$, the volume thermal expansion coefficient is determined as $\beta(P,T) = \frac{1}{V_{eq}(P,T)} \frac{dV_{eq}(P,T)}{dT}\Big|_P$. Additionally, from the information of $\omega_\eta(q, V_i)$ and $V_{eq}(T)$, the phonon dispersions are interpolated on an equilibrium lattice constant at finite T.

The second-order derivative of F(V, T) with T and V is used to determine the heat capacities and bulk moduli. Heat capacity at constant volume (isochoric heat capacity) is $C_V(T) = -T \frac{d^2 F(V,T)}{dT^2}$ and the isothermal bulk modulus is $B_T(V,T) = V_{eq}(T) \frac{d^2 F(V,T)}{dV^2}$. Using the thermodynamic relationships, the heat capacity at constant pressure (isobaric heat capacity) and heat capacity at constant pressure (isobaric heat capacity) are obtained and are given as:

$$C_P(T) = C_V(V,T) + \beta^2(P,T) \cdot V_{eq}(P,T) \cdot B_T(V,T) \cdot T. \tag{4}$$

$$B_S(T) = B_T(T) + \frac{\beta^2(P,T) \cdot B_T^2(V,T) \cdot V_{eq}(P,T) \cdot T}{C_V(V,T)} \tag{5}$$

The contribution of mode η of wave vector **q** to the heat capacity $C_\eta(\boldsymbol{q})$ and the thermal expansion coefficient β is related as:

$$\beta = \frac{1}{B_T} \sum_{\boldsymbol{q},\eta} C_\eta(\boldsymbol{q}) \cdot \gamma_\eta(\boldsymbol{q}) \tag{6}$$

where $B_T$ is the isothermal bulk modulus and $\gamma_\eta(\boldsymbol{q})$ is the mode-Grüneisen parameter, which represents the change in phonon frequency with the volume change, and is represented as: $\gamma_\eta(\boldsymbol{q}) = -\frac{V}{\omega_\eta(\boldsymbol{q})} \frac{d\omega_\eta(\boldsymbol{q})}{dV}$. The mode resolved heat capacity is given as: $C_\eta(\boldsymbol{q}) = \frac{\hbar\omega_\eta(\boldsymbol{q})}{V} \frac{\partial n_\eta(\boldsymbol{q})}{\partial T}$ where $n_\eta(\boldsymbol{q})$ is the Bose-Einstein occupation number, given as $n_\eta(q) = \frac{1}{exp(\frac{\hbar\omega_\eta(q)}{k_BT}) - 1}$. The macroscopic or thermodynamic average Grüneisen parameter γ is related to the microscopic mode-Grüneisen parameter $\gamma_\eta(\boldsymbol{q})$ and $C_\eta(\boldsymbol{q})$ through the relation:

$$\gamma(P,T) = \frac{\sum_{\boldsymbol{q},\eta} C_\eta(\boldsymbol{q}) \cdot \gamma_\eta(\boldsymbol{q})}{\sum_{\boldsymbol{q},\eta} C_\eta(\boldsymbol{q})} = \frac{\beta(P,T) \cdot B_T(V,T) \cdot V_{eq}(P,T)}{C_V(V,T)} \tag{7}$$

**S4: Phonon dispersions at experimental lattice constant at 298 K (6.337 a.u.) for LDA, PBEsol, and PBE, with and without SOC.**

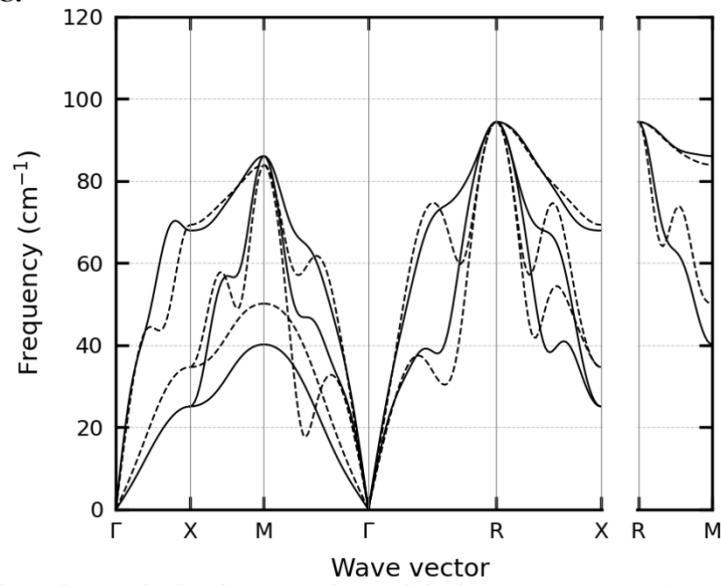

**Fig.S4(A):** Phonon dispersions calculated at experimental lattice constant at 298 K (6.337 a.u.)$^2$ for LDA considering with (solid lines) and without (broken lines) spin-orbit coupling.

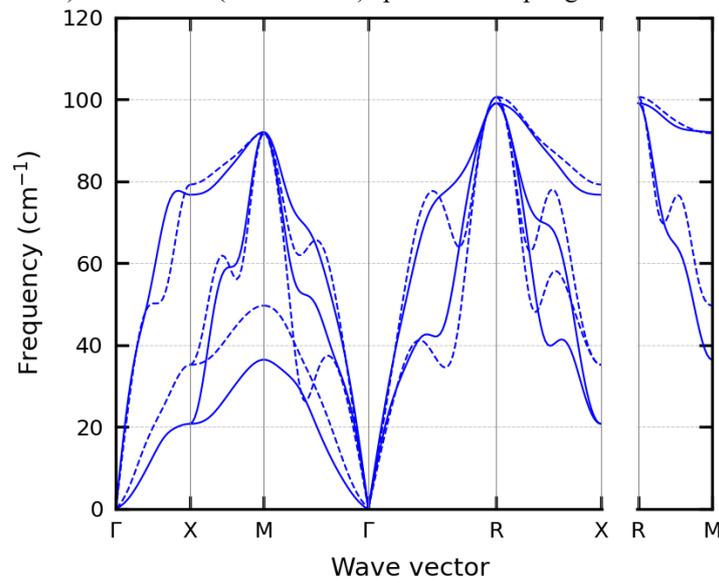

**Fig.S4(B):** Phonon dispersions calculated at experimental lattice constant at 298 K (6.337 a.u.)$^2$ for PBE considering with (solid lines) and without (broken lines) spin-orbit coupling.

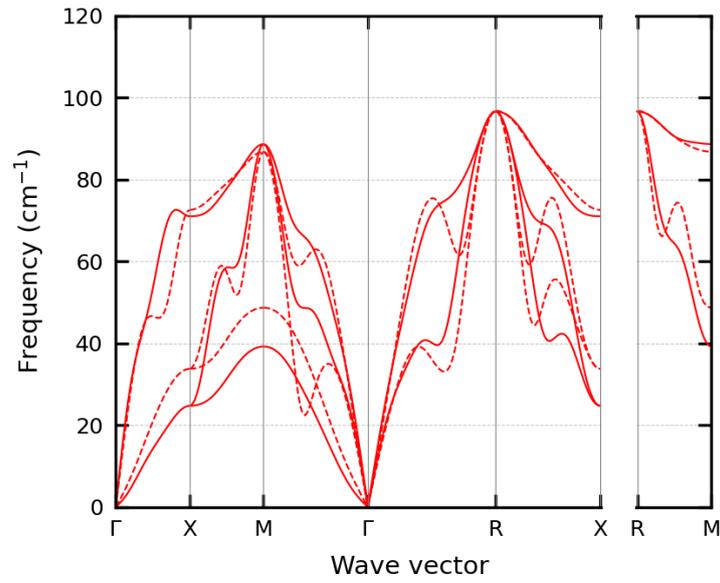

**Fig.S4(C):** Phonon dispersions calculated at experimental lattice constant at 298 K (6.337 a.u.)$^2$ for PBEsol considering with (solid lines) and without (broken lines) spin-orbit coupling.

**S5: Band structure and density of states (DOS):**

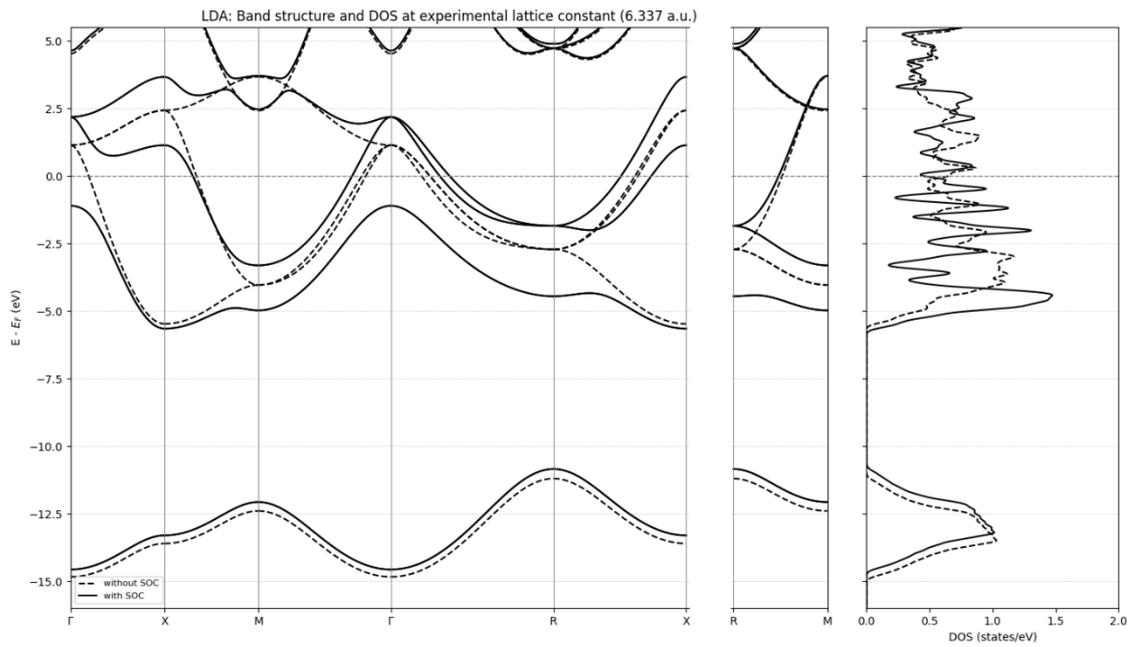

**Fig. S5(A):** Band structure (left) and density of states DOS (right) for LDA calculated with and without spin-orbit coupling (SOC) at experimental lattice constant.

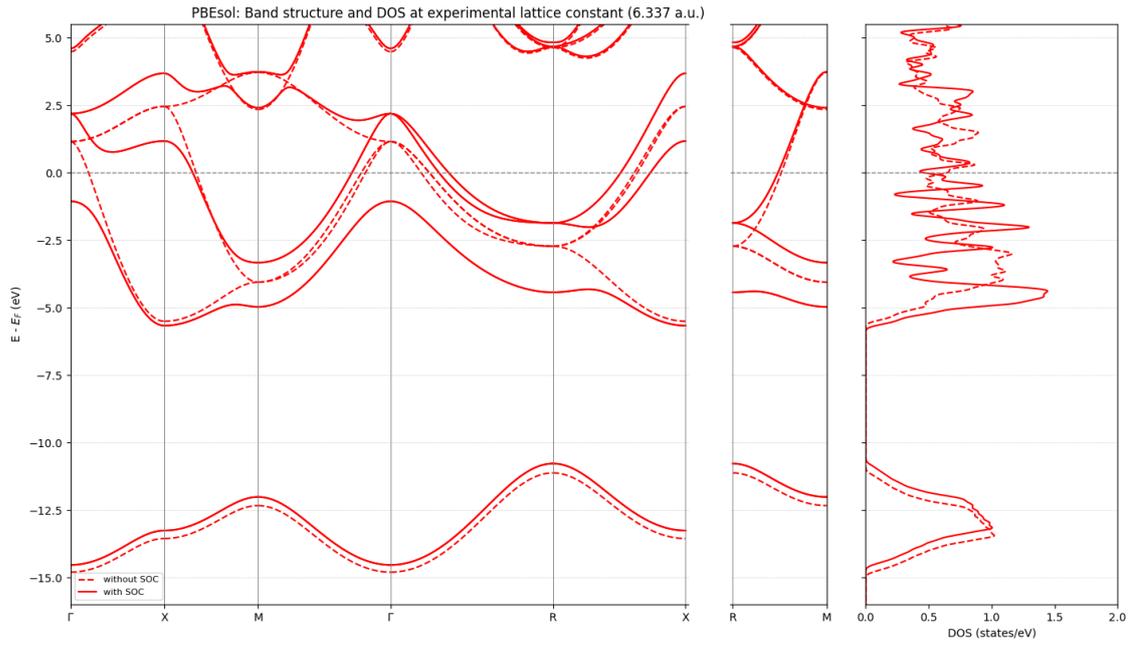

**Fig. S5(B)**: Band structure (left) and density of states DOS (right) for PBEsol calculated with and without spin-orbit coupling (SOC) at experimental lattice constant.

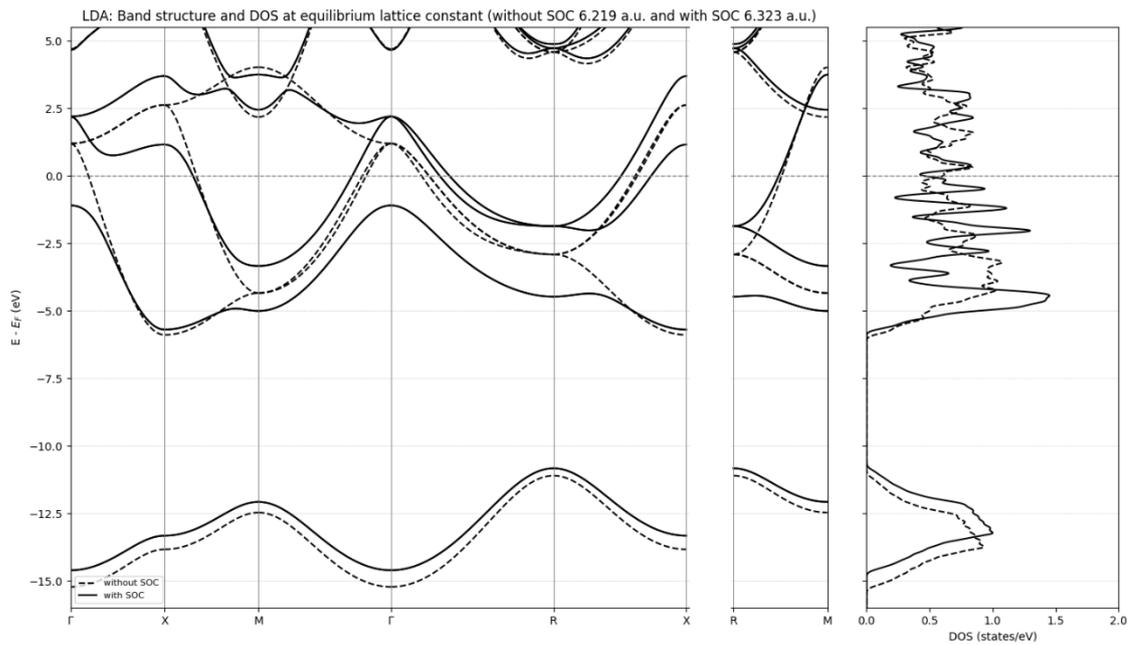

**Fig.S5(C):** Band structure (left) and density of states DOS (right) for LDA calculated with and without spin-orbit coupling (SOC) at the equilibrium lattice constant at 301 K.

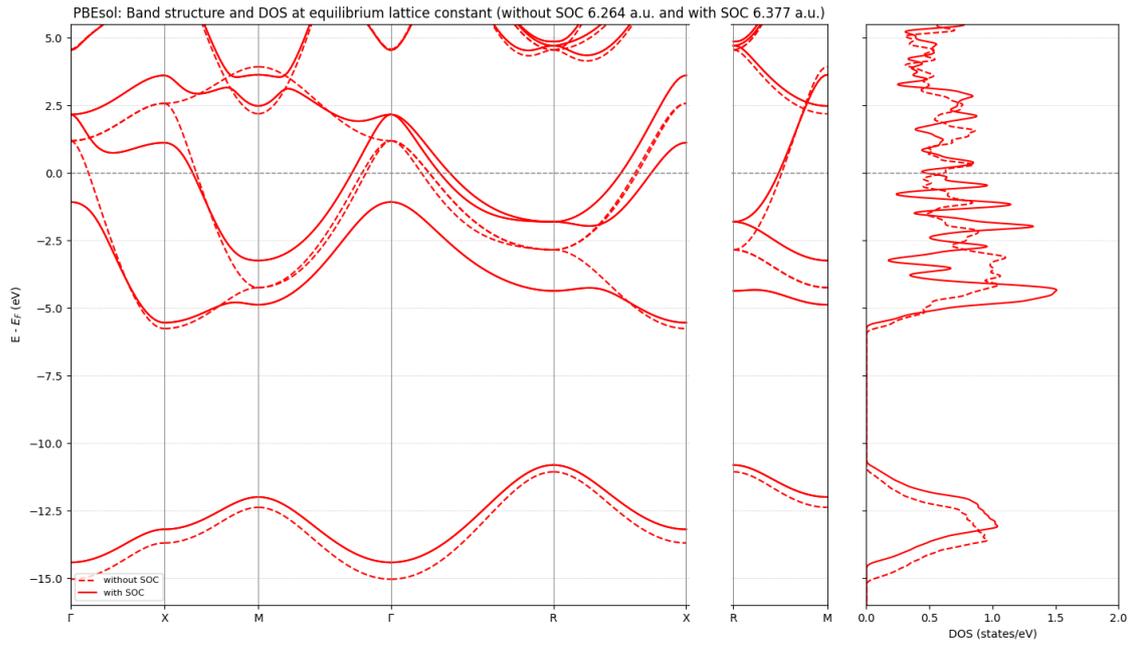

**Fig. S5(D):** Band structure (left) and density of states DOS (right) for PBEsol calculated with and without spin-orbit coupling (SOC) at the equilibrium lattice constant at 301 K.

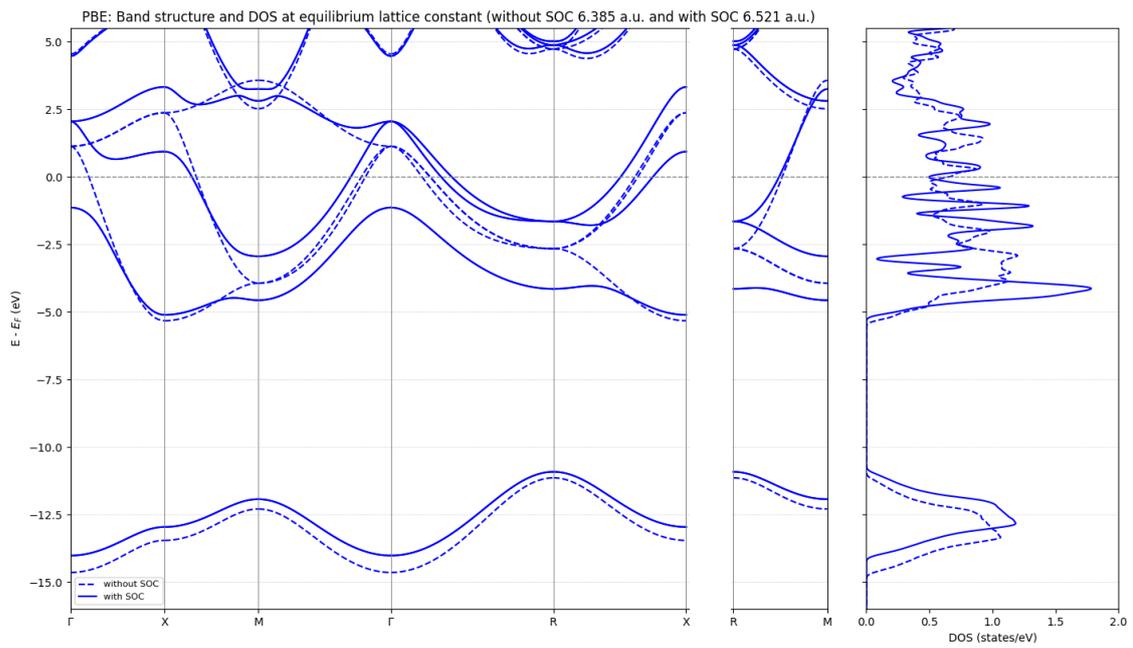

**Fig. S5(E):** Band structure (left) and density of states DOS (right) for PBE calculated with and without spin-orbit coupling (SOC) at the equilibrium lattice constant at 301 K.

**Table S5(A):** Calculated electronic density of states at the Fermi level N(E$_F$), for SC polonium with and without spin orbit coupling (SOC), using LDA, PBEsol, and PBE functionals at both experimental (6.337 a.u.).

| Functional | Pseudopotential | Lattice constant (a.u.) | N(E$_F$) (states/eV) | Decrease in N(E$_F$) with SOC (in %) |
|---|---|---|---|---|
| LDA | With SOC | 6.337 | 0.4582 | 29.5 |
| | Without SOC | 6.337 | 0.6496 | |
| PBEsol | With SOC | 6.337 | 0.4683 | 27.2 |
| | Without SOC | 6.337 | 0.6432 | |
| PBE | With SOC | 6.337 | 0.4797 | 25.2 |
| | Without SOC | 6.337 | 0.6416 | |

**Table S5(B):** Calculated electronic density of states at the Fermi level N(E$_F$), for SC polonium with and without spin orbit coupling (SOC), using LDA, PBEsol, and PBE functionals at equilibrium lattice constants at 301 K.

| Functional | Pseudopotential | Lattice constant (a.u.) | N(E$_F$) (states/eV) | Decrease in N(E$_F$) with SOC (in %) |
|---|---|---|---|---|
| LDA | With SOC | 6.323 | 0.4566 | 24.6 |
| | Without SOC | 6.219 | 0.6058 | |
| PBEsol | With SOC | 6.377 | 0.4691 | 24.0 |
| | Without SOC | 6.264 | 0.6173 | |
| PBE | With SOC | 6.521 | 0.4985 | 24.0 |
| | Without SOC | 6.385 | 0.6559 | |

To quantify the influence of SOC on the electronic structure, we computed the density of states (DOS). We extracted the value of N(E$_F$) for all functionals at both the experimental lattice constants (Table S5A) and the equilibrium lattice constants (Table S5A). For both lattice constant cases, the SOC reduces N(E$_F$) by approximately 25-30% and is functionally independent. This reduction originates from the SOC-induced splitting of the 6*p* bands near the Fermi energy, which broadens the DOS and diminishes sharp features associated with Fermi-surface nesting. Consequently, the reduced N(E$_F$) correlates with the observed weakening of phonon anomalies and the stabilisation of the α-Po phase in the fully relativistic calculations.

.

**S6: Fermi surface nesting for PBE, with and without spin-orbit coupling (SOC):**

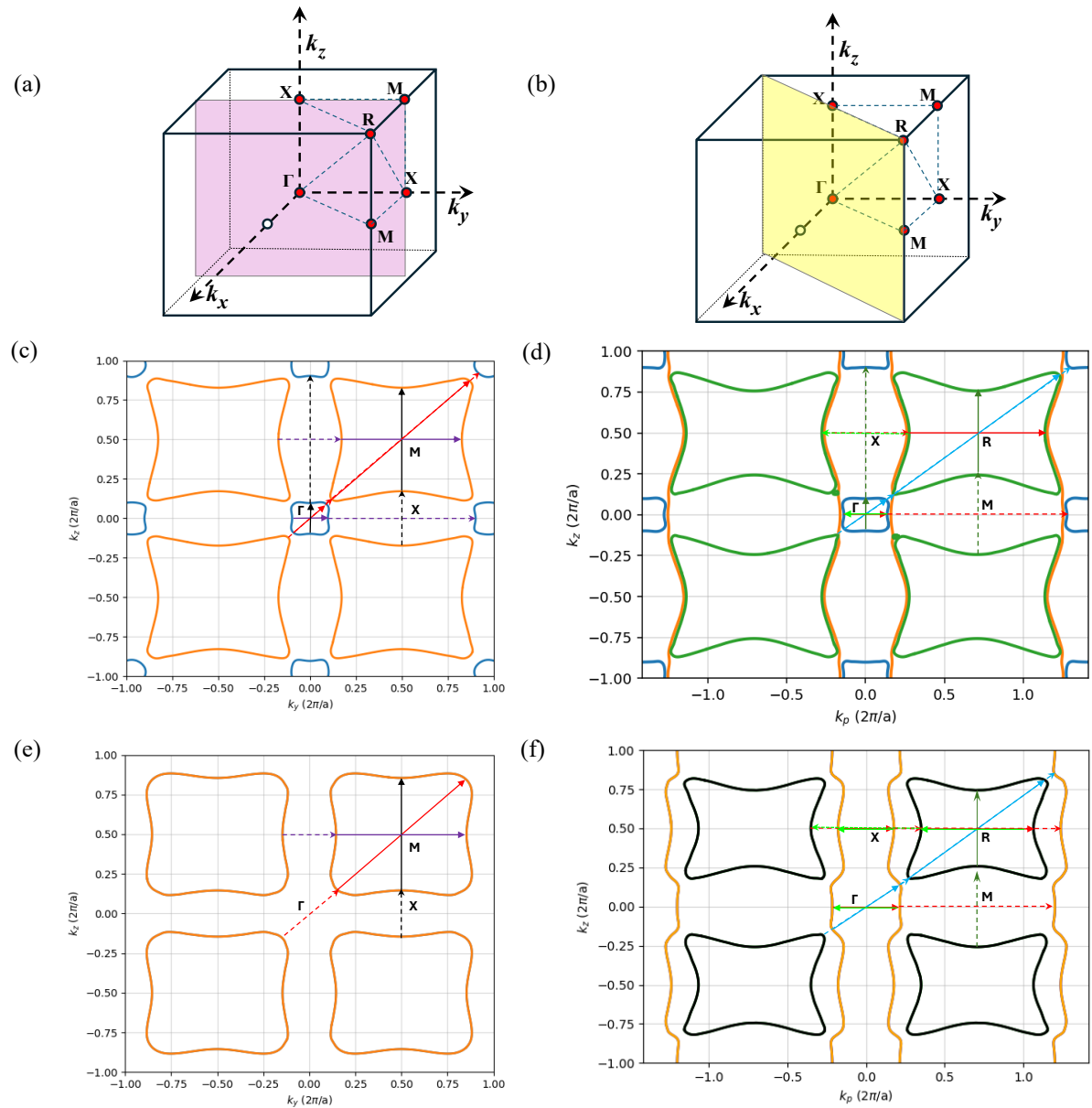

**Fig.S6(A):** (a, b) (200) and ($1\bar{1}0$) planes of the Brillouin zone of simple cubic (sc) polonium. (c and d) show the SR and FR Fermi-surface contour in (200) plane, respectively, while (d and f) show the corresponding SR and FR plots in the ($1\bar{1}0$) plane. All calculations use PBE at the equilibrium lattice constant corresponding to 301 K. Nesting vectors **q** along **Γ - X**, **X - M**, **M - Γ**, **Γ - R**, **R - X** and **M - R** are indicated by purple, black, red, blue, green and olive arrows, respectively.

.

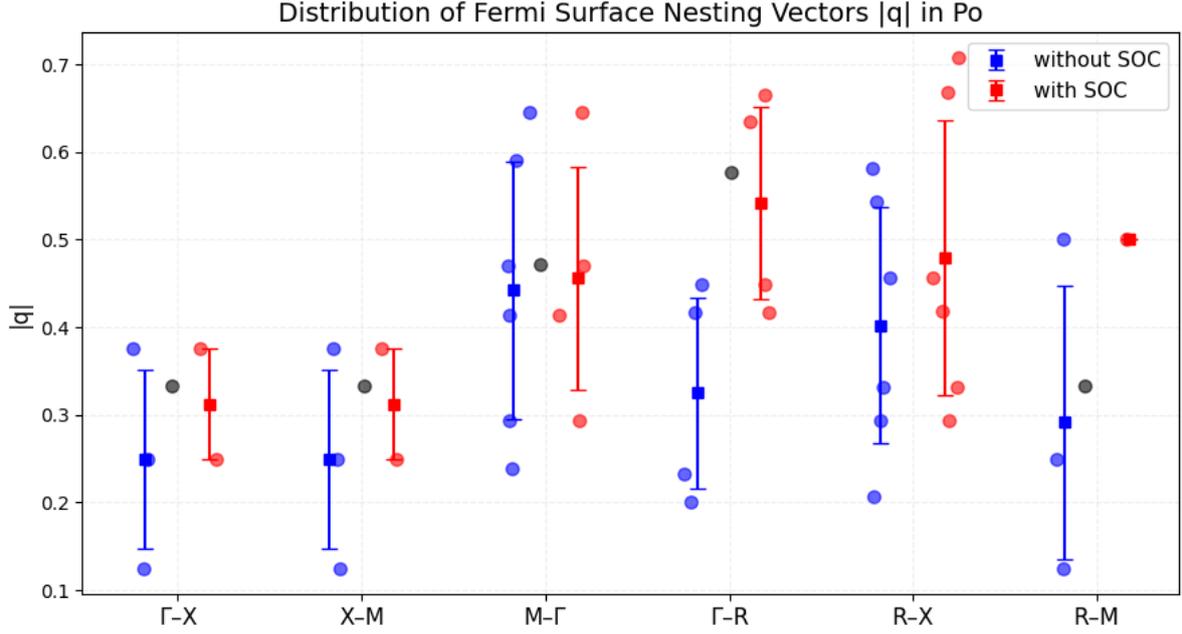

**Fig.S6(B):** Distribution of the magnitude of Fermi-surface nesting vectors |**q**| along the main high-symmetry directions extracted through Fermi-surface matching analysis shown in **Fig.S6(A)**. Blue circles correspond to SR (without SOC) calculations and red markers to FR (with SOC) calculations. Each scatter point represents a nesting vector connecting nearly parallel Fermi-surface segments. Square markers indicate the mean |**q**| with the bar showing the standard deviations. Black markers represent the nesting vectors magnitude reported by Min *et al.*[3] and Belabbes *et al.* [4].

In Fig. S6(A), the contour plot for the SR and FR Fermi surfaces in the (2 0 0) and (1 $\bar{1}$ 0) planes are compared. Nesting vectors were identified along each high symmetry direction where the Fermi surface folds back to the adjacent sections connected by a reciprocal lattice vector **G** i.e., |**q**|=|**q**+**G**|, using a tolerance of |Δ**q**| = 0.1) in the unit of 2π/a). The distribution of |**q**| shown in Fig.S6(B) indicates that the presence of SOC reduces the frequency of nesting vectors along specific directions. Min *et al.*[3] and Belabbes *et al.*[4] reported the nesting vector for the FR case with $\mathbf{q} = \frac{2}{3}$ along **Γ** - **X**, **X** - **M**, **M** - **Γ**, **Γ** - **R**, and **R** - **M** directions. Our results show overall reasonable agreement with Refs.[3,4], except for the **R-M** direction, where the magnitude of **q** differs. We believe this difference is due to the choice of points on the adjacent Fermi surface considered here to study nesting along the **R-M** direction.

**S7: Mode-resolved contribution to the thermal expansion:**

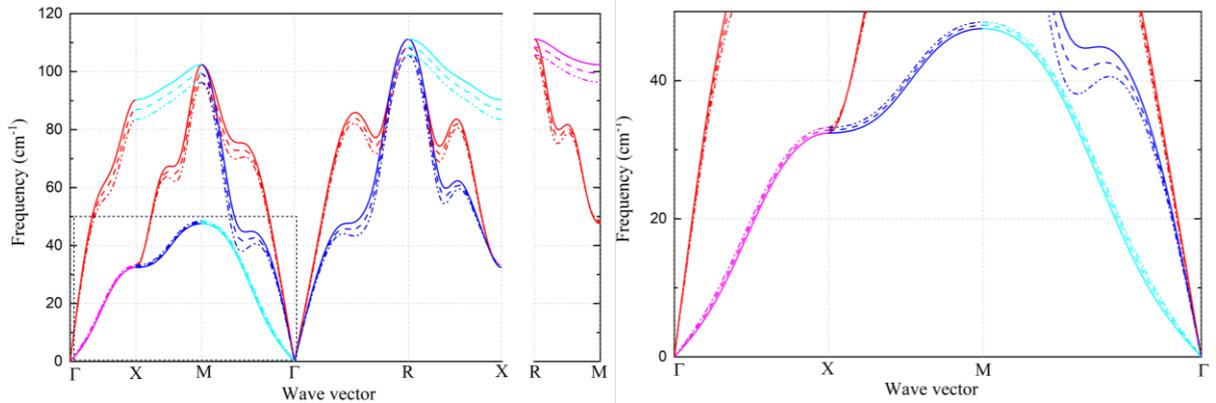

**Fig.S7(A):** (Left) Phonon dispersions for LDA without SOC. (solid: 6.180 a.u., dash: 6.205 a.u., dash-dot-dot: 6.230 a.u.). (Right) Magnified region of the left image showing phonon frequency increases with increasing lattice constant for transverse modes.

To elucidate the microscopic origin of the phonon anomalies, we analyzed the Fermi-surface topology and corresponding nesting vectors of polonium, as illustrated in Fig. S6. In the scalar-relativistic (SR) case, the Fermi surface exhibits pronounced parallel segments connected by nesting vectors primarily oriented along the **Γ-X** direction. These nesting features are known to enhance electron-phonon coupling, resulting in characteristic softening of the transverse phonon branches along the same symmetry lines. Upon inclusion of SOC, the *6p* band splits near the Fermi energy, causing a substantial reduction in the density of states at $E_F$ (~24 %) and a weakening of Fermi-surface nesting. This effect is directly visible in Fig. S5(f), where the spread in the nesting vector along certain directions reduces. The weakening of these electronic instabilities is believed to be observed in the reduction of phonon softening and the disappearance of anomalous kinks in the mode-Grüneisen parameters when SOC is included.

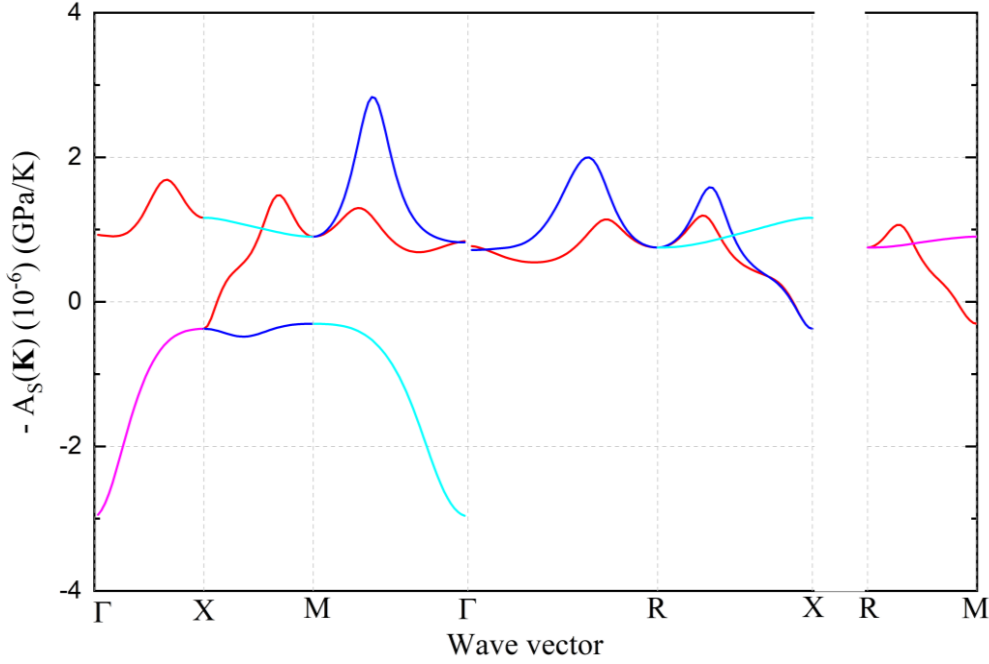

**Fig.S7(B):** The contribution of the phonon modes in the product of volume thermal expansion coefficient and bulk modulus calculated from the phonon frequency for the LDA without SOC case using the phonon dispersion at three lattice constants (6.180 a.u., 6.205 a.u., and 6.230 a.u.).

Following equation 25.15 of Ashcroft and Mermin[5], the contribution of phonon frequency to the volume thermal expansion coefficient is given by:

$$\alpha = \frac{1}{B}\sum_{ks}\left(-\frac{\partial}{\partial V}\hbar\omega_{ks}\right)\frac{\partial}{\partial T}(n_{ks})$$

where the Bose-Einstein occupation $n_{ks} = (e^{\frac{\hbar\omega_{ks}}{k_B T}} - 1)^{-1}$

After carrying out the temperature derivative, the above expression can be written as:

$$\alpha \cdot B = \sum_{ks} A_s(\mathbf{k}), \text{ where } A_s(\mathbf{k}) = -\sum_{ks}\left(\frac{k_B}{\omega_{ks}}\right)\cdot\left(\frac{(\frac{\hbar\omega_{ks}}{k_B T})^2}{(e^{\frac{\hbar\omega_{ks}}{k_B T}}-1)^2}\right)\cdot\left(\frac{\partial\omega_{ks}}{\partial V}\right)$$

Thus, a negative contribution $-A_s(\mathbf{k})$ indicates a mode whose frequency increases with increasing volume and therefore opposes the thermal expansion. Here, we obtained the derivative $\frac{\partial\omega_{ks}}{\partial V}$ using finite differences between the phonon frequencies computed at 6.180 a.u. and 6.230 a.u., while the absolute frequencies from 6.205 a.u. were used as reference. The calculated $-A_s(\mathbf{k})$ values shown in Fig. S7(B), together with the phonon dispersions at the three lattice constants in Fig. S7(A), reveal that some transverse phonon branches harden upon expansion and contribute negatively to $\alpha \cdot B$, thus opposing the lattice expansion.

The results shown in Fig.S7, confirms that the negative $\gamma_{q\eta}$ values originate from specific transverse phonon branches along **Γ-X** and **Γ-M**, whose frequencies increase as the crystal volume expands. This anomalous hardening upon expansion leads directly to negative mode-Grüneisen parameters. In contrast, along **Γ-M** and **Γ-R**, the $\gamma_{q\eta}$ curves exhibit smooth maxima rather than sharp cusps, indicating a continuous anharmonic response without non-analytic behavior except at **Γ**. These maxima primarily result from transverse modes, whose frequency-volume coupling remains strong yet monotonic in nature. Additionally, we find that the inclusion of SOC reduces the magnitude of $\gamma_{q\eta}$ for several branches and modifies the volume dependence of force constants, thereby reshaping the distribution of mode-Grüneisen parameters in a manner consistent with SOC-induced changes to the electronic structure.

## S8. Elastic constant coefficient ($C_{ij}$) and Elastic anisotropic factor(A) as a function of pressure:

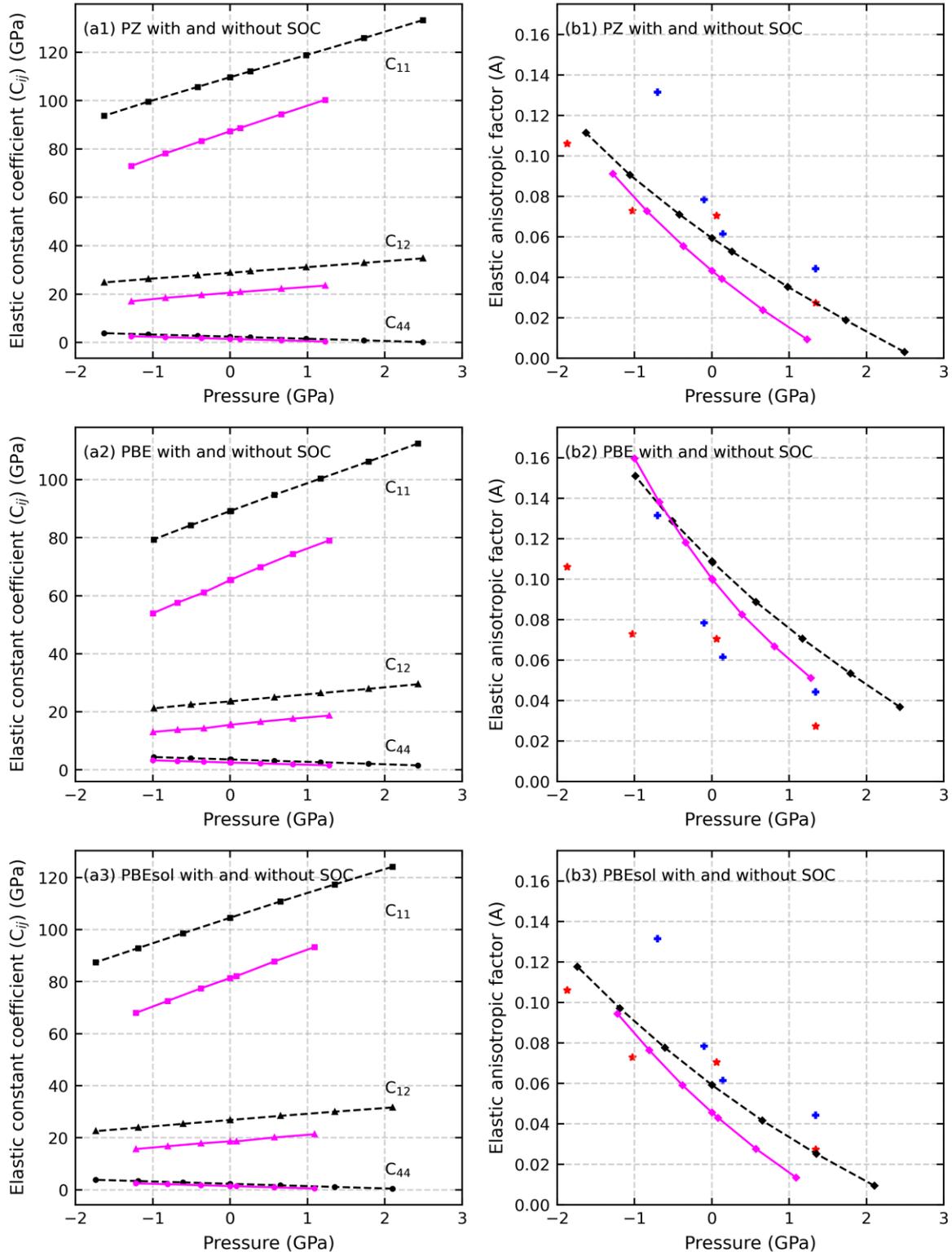

**Fig. S8:** Pressure-dependent (a's) elastic constant coefficient ($C_{ij}$'s) and (b's) elastic anisotropic factor for (1) LDA-PZ, (2) PBE, and (3) PBEsol at 0 K. The solid and broken lines represent the results calculated with and without SOC. The square, triangle, and circle symbols in (a's) are for $C_{11}$, $C_{12}$, and $C_{44}$. In (b's), the diamonds represent the elastic anisotropic factor determined in the present, and the symbols ∗ and + are LDA and LDA+SO results from Ref.[6].

**References:**

1  S. Baroni, S. de Gironcoli, A. Dal Corso and P. Giannozzi, Phonons and related crystal properties from density-functional perturbation theory, *Rev. Mod. Phys.*, 2001, **73**, 515–562.

2  David R. Lide, Ed., *CRC Handbook of Chemistry and Physics*, CRC Press Inc, Boca Raton, Florida, 69th edn., 2005.

3  B. I. Min, J. H. Shim, M. S. Park, K. Kim, S. K. Kwon and S. J. Youn, Origin of the stabilized simple-cubic structure in polonium: Spin-orbit interaction versus Peierls instability, *Phys. Rev. B*, 2006, **73**, 132102.

4  A. Belabbes, A. Zaoui and M. Ferhat, Strong phonon anomalies and Fermi surface nesting of simple cubic Polonium, *Solid State Commun.*, 2010, **150**, 2337–2340.

5  N. W. Ashcroft and N. D. Mermin, *Solid State Physics*, Saunders College Publishing, New York, College Ed., 1976.

6  D. Legut, M. Friák and M. Šob, Why Is Polonium Simple Cubic and So Highly Anisotropic?, *Phys. Rev. Lett.*, 2007, **99**, 016402.